\documentclass[journal=acsnano,manuscript=article]{achemso}
\usepackage[version=3]{mhchem} 

\newcommand{\beginsupplement}{%
        \setcounter{table}{0}
        \renewcommand{\thetable}{S\arabic{table}}%
        \setcounter{figure}{0}
        \renewcommand{\thefigure}{S\arabic{figure}}%
     }

\author{Rocio Camacho-Morales}
\affiliation{ARC Centre of Excellence for Transformative Meta-Optical Systems (TMOS), Department of Electronic Materials Engineering, Research School of Physics, The Australian National University, Canberra, ACT 2601, Australia}
\email{rocio.camacho@anu.edu.au}

\author{Davide Rocco}
\affiliation{Department of Information Engineering, University of Brescia, Via Branze 38, 25123 Brescia, Italy}

\author{Lei Xu}
\affiliation{ARC Centre of Excellence for Transformative Meta-Optical Systems (TMOS), Department of Electronic Materials Engineering, Research School of Physics, The Australian National University, Canberra, ACT 2601, Australia}
\alsoaffiliation{School of Engineering and Information Technology, University of New South Wales, Canberra, ACT 2600, Australia}
\alsoaffiliation{Advanced Optics and Photonics Laboratory, Department of Engineering, School of Science and Technology, Nottingham Trent University, Nottingham, NG11 8NS, UK}

\author{Valerio Flavio Gili}
\affiliation{Matériaux et Phénomènes Quantiques, Université Paris Diderot, F-75013 Paris, France}
\alsoaffiliation{Institute of Applied Physics, Abbe Center of Photonics, Friedrich Schiller University Jena, 07745 Jena, Germany}

\author{Nikolay Dimitrov}
\affiliation{Department of Quantum Electronics, Faculty of Physics, Sofia University, 5 J. Bourchier Blvd., Sofia 1164, Bulgaria}

\author{Lyubomir Stoyanov}
\affiliation{Department of Quantum Electronics, Faculty of Physics, Sofia University, 5 J. Bourchier Blvd., Sofia 1164, Bulgaria}

\author{Zhonghua Ma}
\affiliation{ARC Centre of Excellence for Transformative Meta-Optical Systems (TMOS), Department of Electronic Materials Engineering, Research School of Physics, The Australian National University, Canberra, ACT 2601, Australia}

\author{Andrei Komar}
\affiliation{ARC Centre of Excellence for Transformative Meta-Optical Systems (TMOS), Department of Electronic Materials Engineering, Research School of Physics, The Australian National University, Canberra, ACT 2601, Australia}

\author{Mykhaylo Lysevych}
\affiliation{ARC Centre of Excellence for Transformative Meta-Optical Systems (TMOS), Department of Electronic Materials Engineering, Research School of Physics, The Australian National University, Canberra, ACT 2601, Australia}

\author{Fouad Karouta}
\affiliation{ARC Centre of Excellence for Transformative Meta-Optical Systems (TMOS), Department of Electronic Materials Engineering, Research School of Physics, The Australian National University, Canberra, ACT 2601, Australia}

\author{Alexander Dreischuh}
\affiliation{Department of Quantum Electronics, Faculty of Physics, Sofia University, 5 J. Bourchier Blvd., Sofia 1164, Bulgaria}

\author{Hark Hoe Tan}
\affiliation{ARC Centre of Excellence for Transformative Meta-Optical Systems (TMOS), Department of Electronic Materials Engineering, Research School of Physics, The Australian National University, Canberra, ACT 2601, Australia}

\author{Giuseppe Leo}
\affiliation{Matériaux et Phénomènes Quantiques, Université Paris Diderot, F-75013 Paris, France}

\author{Costantino De Angelis}
\affiliation{Department of Information Engineering, University of Brescia, Via Branze 38, 25123 Brescia, Italy}

\author{Chennupati Jagadish}
\affiliation{ARC Centre of Excellence for Transformative Meta-Optical Systems (TMOS), Department of Electronic Materials Engineering, Research School of Physics, The Australian National University, Canberra, ACT 2601, Australia}

\author{Andrey E. Miroshnichenko}
\affiliation{School of Engineering and Information Technology, University of New South Wales, Canberra, ACT 2600, Australia}

\author{Mohsen Rahmani}
\affiliation{ARC Centre of Excellence for Transformative Meta-Optical Systems (TMOS), Department of Electronic Materials Engineering, Research School of Physics, The Australian National University, Canberra, ACT 2601, Australia}
\alsoaffiliation{Advanced Optics and Photonics Laboratory, Department of Engineering, School of Science and Technology, Nottingham Trent University, Nottingham, NG11 8NS, UK}

\author{Dragomir N. Neshev}
\affiliation{ARC Centre of Excellence for Transformative Meta-Optical Systems (TMOS), Department of Electronic Materials Engineering, Research School of Physics, The Australian National University, Canberra, ACT 2601, Australia}

\title{Infrared up-conversion imaging in nonlinear metasurfaces}

\begin{document}
\newpage
\begin{abstract}
Infrared imaging is a crucial technique in a multitude of applications, including night vision, autonomous vehicles navigation, optical tomography, and food quality control. Conventional infrared imaging technologies, however, require the use of materials like narrow-band gap semiconductors which are sensitive to thermal noise and often require cryogenic cooling. 
Here, we demonstrate a compact all-optical alternative to perform infrared imaging in a metasurface composed of GaAs semiconductor nanoantennas, using a nonlinear wave-mixing process. We experimentally show the up-conversion of short-wave infrared wavelengths via the coherent parametric process of sum-frequency generation. In this process, an infrared image of a target is mixed inside the metasurface with a strong pump beam, translating the image from infrared to the visible in a nanoscale ultra-thin imaging device. 
Our results open up new opportunities for the development of compact infrared imaging devices with applications in infrared vision and life sciences. 
\end{abstract}

\section{Introduction}
Infrared (IR) spectroscopy and imaging are growing in demand due to the increasing number of applications in this spectral region, including optical tomography~\cite{Schmitt:98}, process monitoring~\cite{BOONE2018601}, food and agriculture quality control~\cite{Wang:07}, night vision devices~\cite{Kallhammer}, as well as LIDAR and remote sensing~\cite{Xia:15,Hogstedt:16}. Commercial IR imaging detectors rely on the absorption of incident photons in narrow band gap materials and the release of electrons that are electrically detected. However, due to the low IR photon energy, such IR detection schemes require low-temperature and even cryogenic cooling. As a result, IR cameras are generally bulky, containing several components for photon-electron conversion. 

An alternative scheme, which can potentially overcome the limitations of photoconductive detectors, is the use of nonlinear optical processes for up-conversion of the energy of photons. In this approach, the IR image is not detected directly, instead a parametric nonlinear optical process is employed to convert the image to higher frequencies and detect it using regular cameras, in a process known as up-conversion IR imaging. In 1968, Midwinter first demonstrated IR up-conversion imaging by converting the IR signal to the visible spectrum using a nonlinear crystal and the aid of a pump beam~\cite{Midwinter}. In his work, the spatial information of a short-wave IR (SWIR) image ($\lambda=1.6~\mu$m) was coherently transferred to the visible domain ($\lambda=0.484~\mu$m), using a parametric second-order nonlinear process known as sum-frequency generation (SFG). In the SFG process, two incident waves with frequencies $\omega_1$ and $\omega_2$ interact inside a second-order nonlinear material leading to sum-frequency emission with frequency $\omega_{SFG}$ = $\omega_1$ + $\omega_2$, as shown in Figure~\ref{fgr:Schematic}a. The results obtained by Midwinter showed the possibility of detecting IR images with relatively high sensitivities using standard, fast and uncooled Si-based detectors. 

A decade of intensive research followed the first demonstration of up-conversion IR imaging, where the performance and resolution of the imaging systems were studied in different arrangements, including various incidence angles~\cite{Midwinter:69}, nonlinear bulk crystals~\cite{Andrews}, optical configurations~\cite{Weller}, bandwidth~\cite{Voronin} of the IR radiation, and pump beams~\cite{Abbas:76}. However, the low signal to noise ratio of CCD detectors and poor quality of nonlinear crystals at the time prevented practical developments of up-conversion imaging systems. Recently, the interest in such imaging systems has been renewed, driven by the availability of periodically-poled nonlinear crystals~\cite{Vasilyev:12,Torregrosa:15,Demur}, new laser sources~\cite{Vaughan:11,Huot,Ashik} and the use of intra-cavity configurations~\cite{Jacobo,Torregrosa:15}, which can improve the performance of the systems. The main difficulty in the realization of these IR imaging systems is the phase-matching condition, which not only restricts the conversion efficiency of the up-conversion process but also limits the spectral bandwidth of the IR image, resolution and field of view.

Here, we propose a novel approach to perform IR up-conversion imaging using for the first time, to the best of our knowledge, nanostructured ultra-thin metasurfaces. Our metasurfaces, composed of fabricated nanoantennas on (110) GaAs wafers, are resonant at all the interacting wavelengths. Thus by employing SFG within the resonant metasurface, we demonstrate nonlinear wave-mixing of a SWIR signal beam with a near-IR pump beam, to generate an up-converted emission in the visible spectrum. More importantly, when the IR signal beam carries the image of a target, the spatial information of the target is preserved in the nonlinear wave-mixing process despite being generated by hundreds of independent GaAs crystalline nanoantennas. Therefore, the ultra-fast nonlinear up-conversion process enables IR imaging with femtosecond temporal resolution. Such advancement opens up future opportunities for ultra-fast imaging of chemical reactions in a conventional microscope device.

\section{Nonlinear metasurfaces for IR imaging}

\label{sec:metasurfaces}
Metasurfaces are planar arrays of densely packed nanoantennas designed to manipulate the properties of incident light, including its amplitude, directionality, phase, polarization and frequency~\cite{Neshev:18}.
The optical response of metasurfaces is governed by the collective scattering of individual nanoantennas and the mutual coupling among neighboring nanoantennas. Recent advances in nanofabrication technologies~\cite{Rahmani:18} have motivated extensive research in the field of metasurfaces. Among various examples, dielectric and semiconductor metasurfaces have shown great promise for enhancing nonlinear optical processes at the nanoscale~\cite{deAngelis:2020nonlinear}. Such metasurfaces can exhibit enhanced frequency conversion due to the excitation of optical resonances~\cite{Shcherbakov:2014,Yang:15,Tong:16,Semmlinger:19} and good coupling to free-space. 

However, the strongest nonlinear response of materials originates from quadratic nonlinearity, which is present only in non-centrosymmetric materials. GaAs and its aluminum alloys are often the materials of choice for quadratic nonlinear metasurfaces, being III-V semiconductor materials that possess a zinc-blende non-centrosymmetric crystalline structure, and high quadratic nonlinear susceptibility $\chi^{(2)}\sim200$~pm/V~\cite{Ohashi:1993}. Nevertheless, the use of GaAs compounds comes with significant challenges due to off-diagonal symmetry of its second-order nonlinear susceptibility tensor. While (100)-GaAs metasurfaces have been used to demonstrate ultra-thin second harmonic sources~\cite{Gili:2016:OE, Liu:2016:NL, Camacho2016}, frequency-mixers~\cite{Liu:2018:NATCOMM} and directional lasing~\cite{Ha:18}, harmonic emission from such metasurfaces is forbidden at normal incidence~\cite{Camacho2016, Lochner:2018:ACSPh}. 

Recent studies have aimed at directing the harmonic emission at normal direction to the metasurface~\cite{Lochner:2018:ACSPh,Vabishchevich,Marino:2019:ACSPh,Rocco:2020:IEEE}, however, the most successful strategy to date has been to the use of different crystalline symmetry nanoantennas. Indeed, the GaAs $\chi^{(2)}$ tensor is not invariant under rotation of the crystallographic axes, thus for (111) and (110) GaAs metasurfaces, the diagonal components of the nonlinear susceptibility tensor, $\chi^{(2)}_{\rm rot}$ are different from zero. Normal second harmonic generation (SHG) was first demonstrated in (111) AlGaAs nanoantennas~\cite{Sautter:2019:NL}. However, (110) GaAs nanoantennas have shown highly directional SHG and unique control of its forward to backward emission~\cite{Xu:20}. Such highly directional normal emission promotes the nonlinear mixing of two co-propagating beams to generate sum-frequency emission also propagating along the normal direction. Therefore, in our work we employ (110) GaAs metasurfaces to perform IR up-conversion imaging through the SFG process. In this way, our metasurface can mimic a bulk nonlinear crystal and perform co-linear wave-mixing without the need of co-linear phase-matching.

\begin{figure}[ht!]
\centering
  \includegraphics[width=0.7\textwidth]{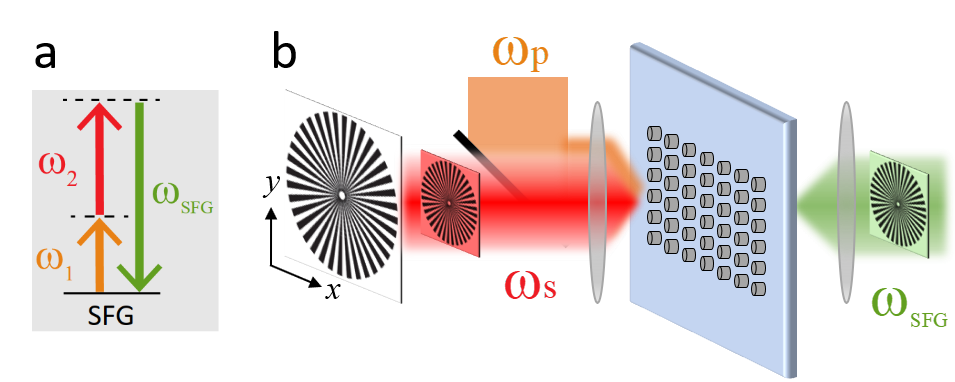}
  \caption{(a) Energy level scheme of sum-frequency generation (SFG) mediated by a second-order nonlinear process. Here, $\omega_1$ and $\omega_2$ are the angular frequencies of the incident signal ($\omega_s$) and pump ($\omega_p$) beams while $\omega_{\rm SFG}$ is the angular frequency of the nonlinear SFG emission. The solid and dashed black lines indicate real and virtual energy levels, respectively. (b) Schematic illustrating the concept of IR nonlinear imaging. The pump beam ($\omega_p$) and the IR signal beam ($\omega_s$), encoding the image of a target (\textit{i.e.} a Siemens star), are simultaneously focused by a lens on a (110) GaAs metasurface. At the output of the metasurface, a visible image of the target is obtained through the SFG emission ($\omega_{\rm SFG}$). In the schematic, the focused pump and signal beams are not spatially overlapped on the metasurface only for visualisation purposes.}
  \label{fgr:Schematic}
\end{figure}

The process of IR up-conversion imaging in a nonlinear metasurface is schematically represented in Figure~\ref{fgr:Schematic}b. In this Figure, the image of a target (Siemens star) is encoded in the IR signal beam (red beam) and up-converted to a visible image (green beam) due to the nonlinear wave-mixing of signal and pump beams (orange beam) within the metasurface. In the rest of the paper, the colors red, orange and green in the figures will be used to refer to the signal, pump and SFG beams, respectively. In our configuration, the pump beam and the IR image of a target in the signal beam, are simultaneously focused on the metasurface (see left-hand side of Figure~\ref{fgr:Schematic}b). The pump and signal beams are mixed together within the GaAs metasurface through the SFG process (see energy diagram in Figure~\ref{fgr:Schematic}a), resulting in up-converted photons that form a visible image of the target (see right-hand side of Figure~\ref{fgr:Schematic}b). 

In our experiments, we chose the signal beam wavelength at 1530~nm. This choice is dictated by applications in night-vision technologies and corresponds to the band of maximum glow of the night sky ($1500 - 1700$~nm) \cite{Krieg:19}. The pump beam is chosen at 860~nm since commercial high-power laser diodes at this wavelength are widely available. Moreover, this wavelength is invisible to human eye. With this choice of pump and signal wavelengths, the SFG frequency-mixing process results in the generation of visible green light (at 550~nm), where the human eye has maximum sensitivity.

\section{Numerical results}
\begin{figure}[ht!]
   \centering
   \includegraphics[width=0.98\textwidth]{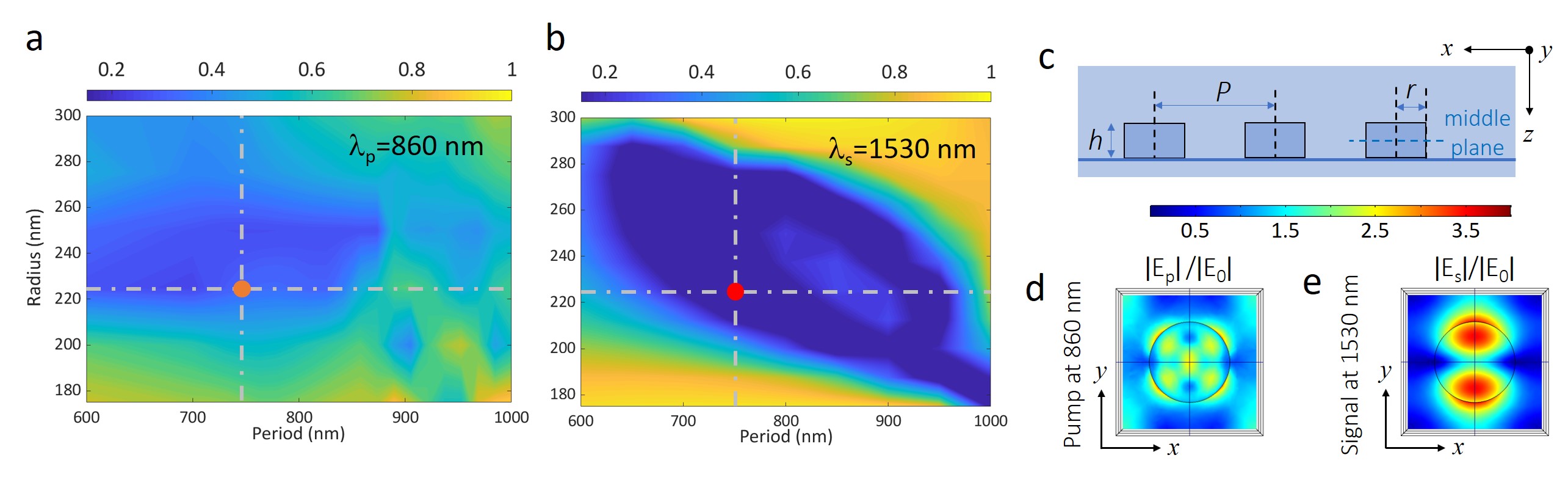}
   \caption{Calculated transmittance of GaAs metasurface as a function of radius and periodicity of the nanoantennas for an incident wavelength of (a) 860 (pump) and (b) 1530~nm (signal). The nanoantenna height is fixed at 400~nm. In each plot, the transmittance is indicated by the top color bar. A simultaneous double resonant behavior is achieved when the periodicity is 750~nm and the radius is 225~nm (see orange and red dots). (c) Side view of the designed metasurface illustrating the height (\textit{h}), radius (\textit{r}), and middle \textit{xy}-plane of the nanoantenna, as well as periodicity (\textit{P}). Calculated modulus of the electric field distribution in a metasurface unit cell ($r=225$~nm, $h=400$~nm) for an incident wavelength of (d) 860 and (e) 1530~nm. The field intensity in both plots is represented by the top color bar. The calculations show the middle \textit{xy}-plane cut view of a nanoantenna.}
   \label{fgr:Calculations}
\end{figure}

First, we designed the linear optical properties of (110) GaAs metasurfaces to support resonances at the wavelengths of the signal and pump beams. 
Different geometric parameters of the metasurface were optimized to obtain the desired resonances, namely the disk nanoantenna radius, $r$ and the array periodicity, $P$ (see Figure~\ref{fgr:Calculations}c). In our calculations, the height of the nanoantennas, $h$ is fixed to 400~nm. The 2D transmission maps obtained by varying the nanoantennas separation from 600 to 1000~nm, and the nanoantennas radius from 175 to 300~nm are shown in Figures~\ref{fgr:Calculations}a and b. Figure~\ref{fgr:Calculations}a was calculated at an incident wavelength of 860~nm, corresponding to the pump beam, while Figure~\ref{fgr:Calculations}b was calculated at the signal wavelength, 1530~nm. In both cases we used normal plane-wave illumination. The areas of interest in these 2D maps are the low transmission areas, indicated by blue colored regions. These transmission dips correspond to a strong resonant behavior of the metasurface. As can be seen in Figure~\ref{fgr:Calculations}a, strong resonances can take place in metasurfaces with periodicity in the range of approximately 600 to 850~nm and radius in the range of approximately 220 to 250~nm. Similarly, Figure~\ref{fgr:Calculations}b shows that strong resonances can be obtained in metasurfaces with periodicity in the range of 650 to 800~nm and radius in the range of approximately 220 to 280~nm, along with other strong resonances observed in metasurfaces with larger periodicity. Thus, these transmission maps offer a range of geometric parameters of metasurfaces where strong resonances are simultaneously obtained for the signal and pump beams. The pair of geometric parameters (\textit{P} and \textit{r}) indicating a double resonant behavior of the metasurface were next used to simulate the SFG emission. The final design consists of a (110)-GaAs metasurface with $P=750$~nm, $r=225$~nm (see orange and red dots in Figure~\ref{fgr:Calculations}a and b, respectively) and $h=400$~nm, where strong double resonant behavior and maximum SFG emission normal to the metasurface (see Supporting Information, Fig.~S4) were obtained. The forward SFG conversion efficiency, $\eta=P_{SFG}/P_s$ of the designed metasurface is $1.6 \times 10^{-6}$ for $I_p=0.78$~GW/cm$^2$ and $I_s=0.38$~GW/cm$^2$, corresponding to the typical values in our measurements. This efficiency is dependent on the pump power, therefore the normalized conversion efficiency $\eta_{Norm} = \eta/P_p$, is a better measure of the efficiency of the SFG process. Here $P_{p}$ is the average power of the pump beam.

According to the final design of the metasurface, the spatial field profiles in a metasurface unit cell were calculated at the pump and signal wavelengths. These field profiles normalized to the incident electric field are shown in Figure~\ref{fgr:Calculations}d and e, respectively. In each case, the electric field profile is shown in the middle \textit{xy}-plane of the nanoantenna (see Figure~\ref{fgr:Calculations}c). The spatial field profile of the pump (Figure~\ref{fgr:Calculations}d) shows a maximum enhancement of about 2.5 times with respect to the incident field. At the edges of the nanoantenna, the spatial distribution of the pump field shows a four-lobe pattern, resembling a quadrupolar resonant mode. Multipolar decomposition of the total scattering indicates the contribution of an electric and magnetic dipole, followed by an electric and magnetic quadrupole at the pump wavelength (see Supporting Information, Fig.~S1). The field spatial profile of the signal beam, illustrated in Figure~\ref{fgr:Calculations}e, shows a field enhancement of more than three with respect to the incident field (see Figure~\ref{fgr:Calculations}b). The spatial distribution of the signal field is indicative of a dipole mode excitation. Multipolar decomposition of the total scattering corroborates the excitation of a strong magnetic dipole, followed by an electric dipole at the signal wavelength (see Supporting Information, Fig.~S1). The electric field profiles of the pump and signal beams in the \textit{xz}- and \textit{yz}-plane of the metasurface are shown in Supporting Information, Fig.~S2.

\begin{figure*}[htb]
   \centering
   \includegraphics[width=0.75\linewidth]{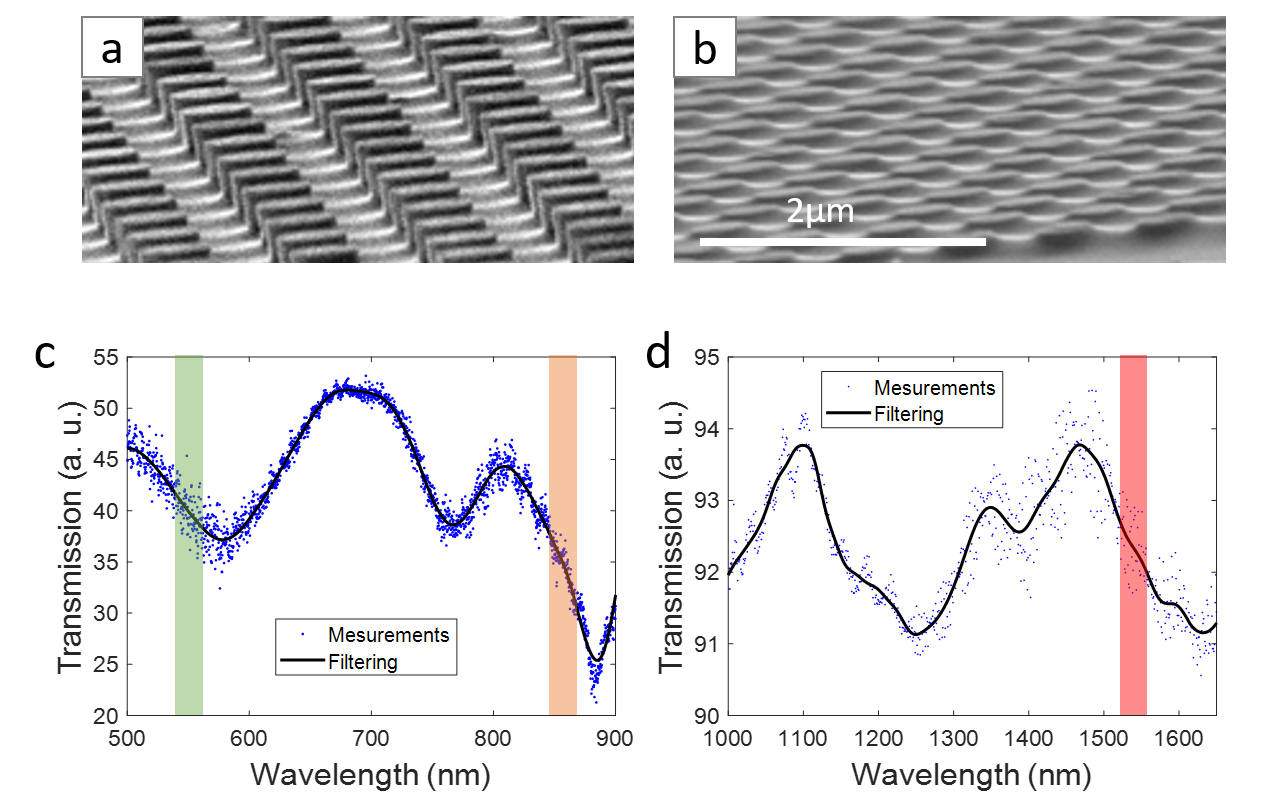}
   \caption{(a,b) Oblique SEM images of a GaAs wafer consisting of the fabricated nanoantennas (a) before and (b) after lift-off from the substrate. Experimental transmission spectra of GaAs metasurface measured in the (c) visible and (d) IR spectral regions. The wavelength positions of the SFG, pump and signal are indicated by the green, orange and red vertical lines, respectively.}
   \label{fgr:Linearspectra_meta}
\end{figure*}

\section{Experimental results}
\subsection{Fabrication and linear characterization of the metasurface}

According to our optimal design, we fabricated metasurfaces with $h=400$~nm, $r=225$~nm and $P=750$~nm (see Figures~\ref{fgr:Linearspectra_meta}a and b). Our fabrication technique allows us to perform IR imaging in a transmission configuration, as typically required in night vision and imaging applications. 
The linear transmission spectrum of the fabricated metasurface was measured by white light spectroscopy. The visible transmission spectrum in Figure~\ref{fgr:Linearspectra_meta}c, shows three strong resonances (manifested as dips) centered around 580, 770 and 885~nm. The vertical orange line indicates the position of the pump wavelength (at 860~nm). In addition, a resonant behavior is observed at the corresponding SFG wavelength (550~nm), indicated by a vertical green line. Figure~\ref{fgr:Linearspectra_meta}d shows the IR transmission spectrum with two strong resonances centered around 1250 and 1630~nm and a small resonance centered around 1385~nm. The vertical red line indicates the position of the signal wavelength (at 1530~nm), where the metasurface exhibits a resonant behavior. The measured transmission spectra of the metasurface (see blue dots in Figures~\ref{fgr:Linearspectra_meta}c and d) exhibit periodic fringes due to Fabry-Perot interference in the substrate. An average filter function from Matlab was used (solid black line in Figures~\ref{fgr:Linearspectra_meta}c and d) to smooth the transmission spectra. Overall, our fabricated GaAs metasurface exhibits a multi-resonant behavior, with resonances around the three wavelengths of interest at the pump, signal and SFG beams. However, the SFG efficiency is not only determined by the far-field scattering spectrum but also by the near-field enhancement and spatial mode-overlap. Therefore, experiments are required to optimize the efficiency of the nonlinear sum-frequency process.

\begin{figure}[ht!]
\centering
\includegraphics[width=0.85\textwidth]{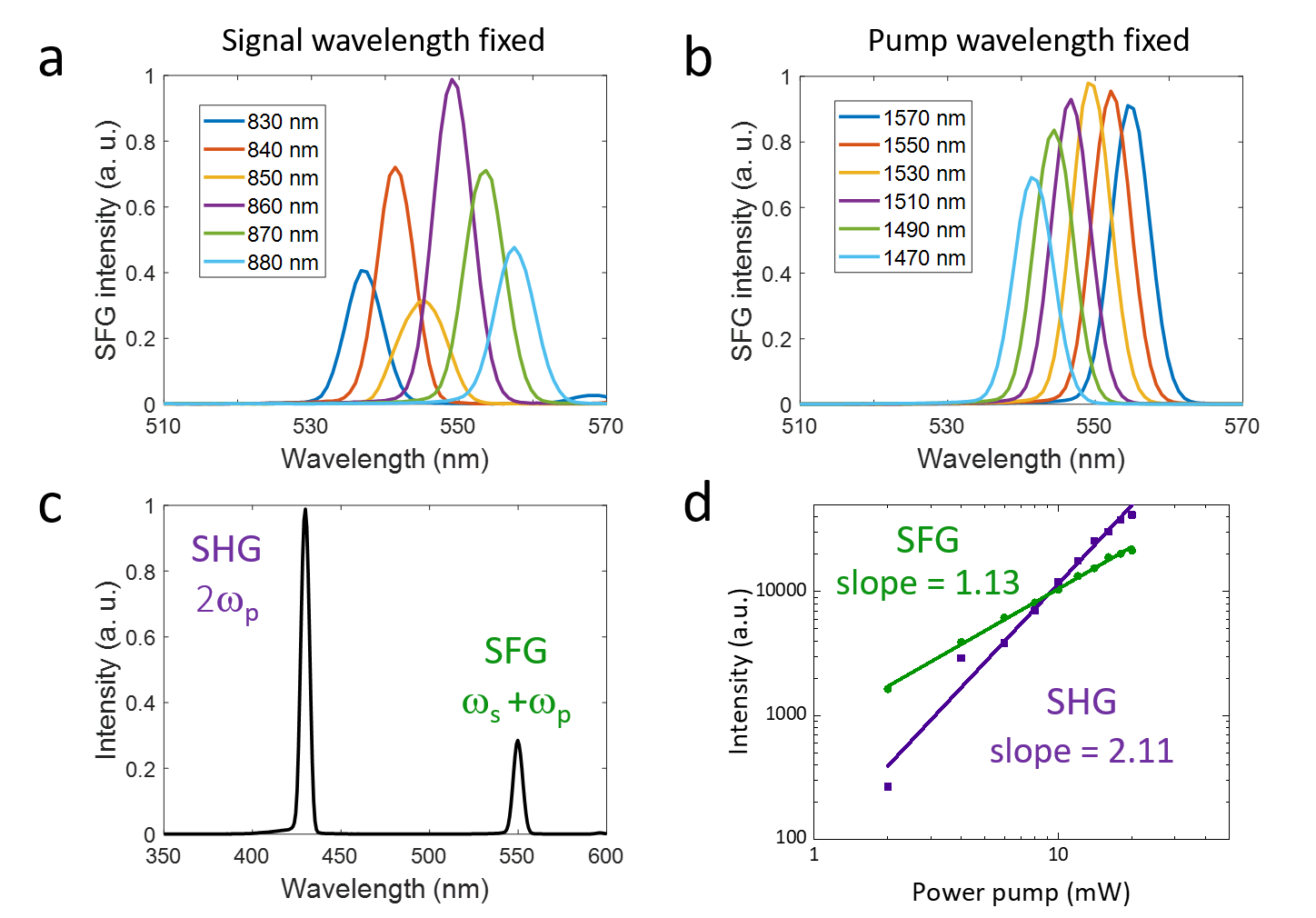}
\caption{Spectral dependence of the SFG emission on the varying wavelength of (a) the pump beam from 830 to 880~nm, and (b) the signal beam from 1470 to 1570~nm. In the former case the wavelength of the signal is fixed at 1530~nm, while in the latter case the wavelength of the pump is fixed at 860~nm. (c) Normalized nonlinear spectrum from metasurface, where two strong emission peaks centered at 430 and 550~nm are generated by the SHG$_{2\omega_p}$ of the pump and the SFG process, respectively. (d) Intensity dependence of SHG$_{2\omega_p}$ and SFG on pump power, shown in a log-log plot. The SHG$_{2\omega_p}$ intensity depends quadratically on the pump power, while the SFG intensity depends linearly.}
\label{fgr:Nonlinear_spectra}
\end{figure}

\subsection{Nonlinear emissions from GaAs metasurfaces}

Next, we measured the SFG intensity by independently tuning the wavelengths of the signal and pump beams around the spectral region of interest. We use a Ti:Sapphire laser with an optical parametric oscillator (OPO) which together deliver two pulsed train beams (see Supporting Information Figs. S5 and S7) with a repetition rate of 80~MHz. First, the wavelength of the signal beam was fixed at 1530~nm and the wavelength of the pump was tuned from 830 to 880~nm (see Figure~\ref{fgr:Nonlinear_spectra}a). Both beams were linearly polarized along the horizontal direction. The spectrum in Figure~\ref{fgr:Nonlinear_spectra}a shows the emission of the SFG from 537 to 558~nm, with a maximum efficiency at 549~nm, corresponding to an excitation pump beam at 860~nm. After exhibiting a maximum at 549 nm, the SFG intensity decreases with the increase of the pump wavelength. The use of a pump beam with a wavelength longer than 880~nm is limited by our laser system.

Next, the wavelength of the signal beam was tuned from 1470 to 1570~nm, while maintaining the pump beam fixed at 860~nm (see Figure~\ref{fgr:Nonlinear_spectra}b). The wavelength of the pump beam was chosen according to the maximum SFG observed in Figure~\ref{fgr:Nonlinear_spectra}a. Figure~\ref{fgr:Nonlinear_spectra}b shows the SFG emission from 541 to 555~nm, with a maximum at 549~nm, corresponding to a signal beam at 1530~nm. After the maximum at 549~nm, the SFG intensity gradually decreases with the increase of signal wavelength. Through all these measurements, the average power of the pump and signal beams, measured right before the metasurface, was kept constant at 18 and 14~mW, respectively. As can be seen in Figures~\ref{fgr:Nonlinear_spectra}a and b, the optimized SFG intensity takes place when the metasurface is excited by a pump beam at 860~nm and a signal beam at 1530~nm. This behavior is explained by the near-field enhancement of the excitation beams, when the metasurface is resonantly excited (see Figure~\ref{fgr:Linearspectra_meta}c and d) at these excitation wavelengths.

The visible nonlinear spectrum of the metasurface is shown in Figure~\ref{fgr:Nonlinear_spectra}c, characterized by two strong nonlinear emissions at 430 and 550~nm. The metasurface was excited by signal and pump beams at the optimized wavelengths using an average power of 10~mW in each beam, measured before the metasurface. The shorter wavelength nonlinear emission at 430~nm originates from the SHG of the pump ($2\omega_p$), while the emission at 550~nm originates from the SFG process ($\omega_s+\omega_p$). Other nonlinear processes are also generated in the metasurface at wavelengths longer than the SFG, however, these wavelengths are blocked by the short-pass filter (with cut-off wavelength of 600~nm) used to filter out the transmitted pump beam (see Supporting Information, Fig.~S5), thus they are not collected by our detection system. 

Above the band gap of the GaAs (1.42~eV, 873~nm), the absorption coefficient increases as the incident wavelength decreases. Therefore, in Figure~\ref{fgr:Nonlinear_spectra}c the higher intensity of the SHG$_{2\omega_p}$, as compared to the SFG intensity, is unexpected. Traditionally, the stronger nonlinear intensity of a metasurface is associated to the near-field enhancement at the fundamental wave~\cite{Liu:2018:NATCOMM,Semmlinger:19}. However, the efficiency of the nonlinear frequency-mixing also depends on the spatial mode overlap of the interacting waves~\cite{Celebrano,Colom,Harutyunyan} (Figure~\ref{fgr:Calculations}d and e). Inside the GaAs nanoantennnas the field enhancement of signal and pump beams has similar intensity (see Supporting Information, Fig.~S2, as this is not reflected in the particular cross-section of Figures~\ref{fgr:Calculations}d and e). Therefore, the higher intensity of the SHG ($\omega_p+\omega_p$) can be attributed to the full spatial overlap of the pump field with itself (Figure~\ref{fgr:Calculations}c). Whereas, in the case of SFG ($\omega_s+\omega_p$) the spatial overlap between the signal and pump beams (Figures~\ref{fgr:Calculations}c and d) is not complete, thereby the intensity of the SFG emission is weaker. It is worth noting that the SFG intensity is also dependent on the spatial overlap between the excitation and SFG fields (see Supporting Information, Fig.~S3). Further studies can be performed to investigate the relative intensities of the nonlinear emissions. However, these studies are outside the scope of our work. 

To verify the origin of the nonlinear emissions generated by the metasurface, the average power of the pump beam was gradually increased from 2 to 20~mW with a 2~mW step, while keeping the power of the signal beam constant. The intensity of the parametric emissions was recorded and analyzed on a log-log plot, as shown in Figure~\ref{fgr:Nonlinear_spectra}d. In the case of SHG$_{2\omega_p}$ a quadratic dependence on the pump power was obtained with a slope of 2.11; whereas in the case of the SFG a linear dependence on the pump power with a slope of 1.13 was obtained. This is expected, since in the SFG process the pump field contributes with a single photon to the frequency up-conversion (Figure~\ref{fgr:Schematic}a). 
The measured SFG conversion efficiency achieved for the metasurface was $\eta=6\times 10^{-8}$ at average beam powers of $P_p^{ave}=16.4$~mW and $P_s^{ave}=16.8$~mW, corresponding to average intensities of $I_p^{ave}=0.8$~GW/cm$^2$ and  $I_s^{ave}=0.4$~W/cm$^2$ (see Supporting Information). This efficiency is lower than our numerical estimation, which can be attributed to fabrication imperfections.

\begin{figure}[tb]
\centering
\includegraphics[width=0.40\textwidth]{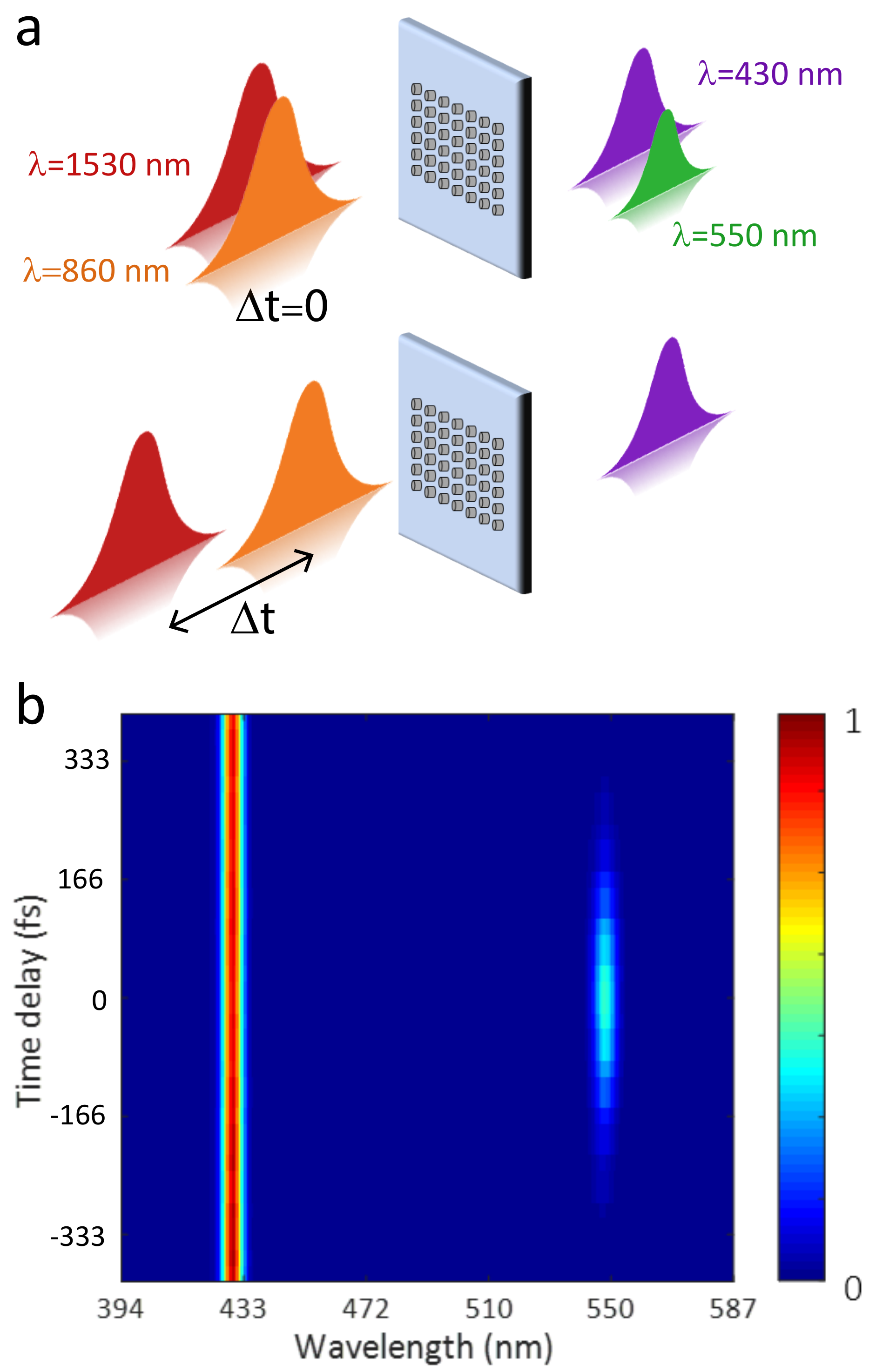}
\caption{(a) Schematic illustration of nonlinear emissions generated by metasurface at different time delays, $\Delta t$ between pump and signal pulses. SHG$_{2\omega_p}$ ($\lambda$=430~nm) and SFG ($\lambda$=550~nm) are generated by the metasurface when the time delay between pulses is zero (top row). The SFG emission vanishes when the time delay increases (bottom row). (b) Two-dimensional nonlinear spectra measured as a function of time delay. The strongest spectral emission centered at 430~nm is independent of the time delay, while the emission centered at 550~nm is strongly dependent of the time delay, having a width of $267$~fs.}
\label{fgr:Delayline}
\end{figure}

As an important consequence, when short pulses are used for both signal and pump beams in the SFG process, these two pulses require temporal synchronization ($\Delta t=0$), as illustrated in the top of Figure~\ref{fgr:Delayline}a. In contrast, the SHG$_{2\omega_p}$ is a degenerated nonlinear process where two photons come from the same pulse, thus no temporal synchronization is required in this case (see bottom of Figure~\ref{fgr:Delayline}a). Here, we achieved temporal synchronization of the signal and pump pulsed beams by using a free-space variable delay line with micrometer adjustment to finely adjust the path length of the pump beam (see Supporting Information, Fig.~S5), and thus to accurately control the time delay between pulses. The spectra of the up-converted nonlinear emissions were measured as a function of the time delay, as illustrated in Figure~\ref{fgr:Delayline}b. The experimental conditions used in these measurements were the same as the ones used in Figure~\ref{fgr:Nonlinear_spectra}c. It is worth noting that the SHG$_{2\omega_p}$ is independent of time delay, while the strongest SFG emission is generated at zero time delay. As can be seen in Figure~\ref{fgr:Delayline}b, when the time delay changes from zero to $\pm333$~fs, the SFG intensity continuously drops until becoming negligible. The temporal duration of the SFG emission was measured to be 267~fs, which is effectively the convolution of the signal and pump pulses (see Supporting Information, Fig.~S7). This finding demonstrates that the up-conversion process preserves the temporal information of the femtosecond IR pulses and can find applications in ultra-fast IR imaging of dynamic phenomena~\cite{Antonucci:12}.

\subsection{Infrared imaging}

\begin{figure}[tb]
\centering
\includegraphics[width=0.8\textwidth]{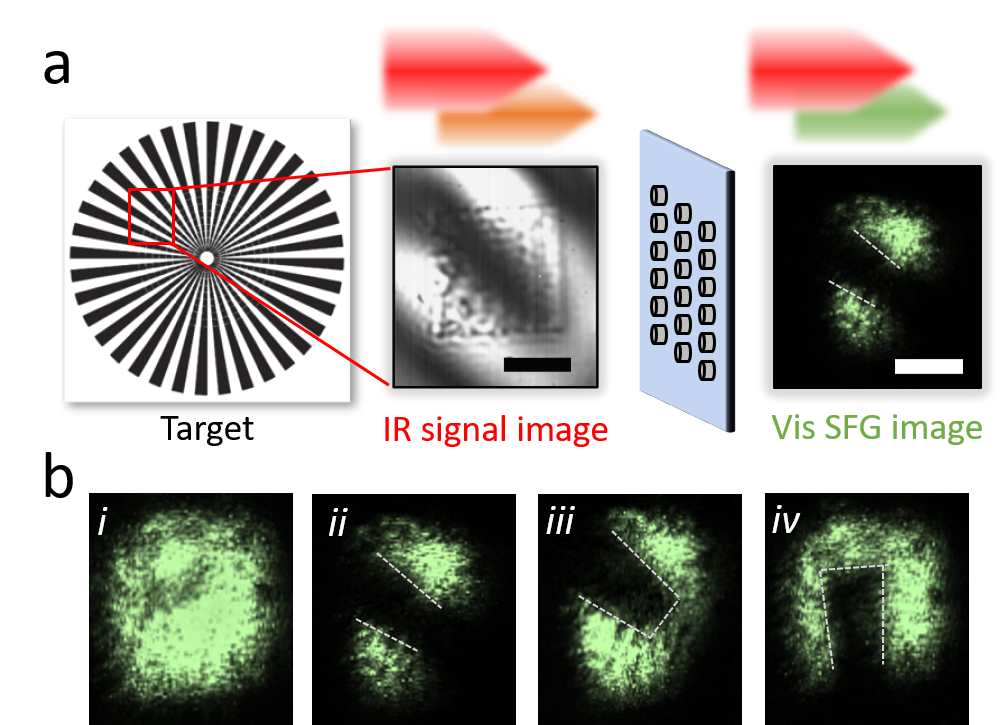}
\caption{(a) IR up-conversion imaging using a GaAs metasurface: the IR signal beam passes through a target, which is imaged onto the metasurface. Mixing with the pump beam results in a visible image of the target (in the SFG beam), which is subsequently imaged by a lens onto a camera. The IR and visible images are acquired with an InGaAs and CMOS cameras, respectively. They correspond to a section of the Siemens star, highlighted by the red square. The scale bars on both images are 15~$\mu$m. (b) Up-converted images at different transverse positions of the target: (i) SFG beam when the target is removed and (ii-iv) SFG images for three positions of the target in the transverse plane.}
  \label{fgr:Vis&IRImages}
\end{figure}

Finally, we present the up-conversion IR imaging enabled by the GaAs metasurface. In this experiment, the collimated signal beam passes through a Siemens star target (see Supporting Information, Fig.~S6), which is imaged by the focusing lens onto the metasurface. The IR image in Figure~\ref{fgr:Vis&IRImages}a (second frame) was acquired with an InGaAs IR camera (Xenics, XS-1.7-320), using only signal beam illumination. The section of the Siemens star imaged is highlighted by a red square. At the metasurface, this SWIR image is mixed with the mildly focused pump beam and through the SFG process converted to a visible wavelength (Figure~\ref{fgr:Vis&IRImages}a, last frame). Importantly, the visible images presented here have been captured using a conventional CCD camera (Starlight Xpress, SXVR-H9). Ideally, since the SFG is a coherent parametric process, all the spatial information in the SWIR image should be preserved to the visible. However, in the case of a metasurface the different spatial features of the target are being converted by hundreds of independent nanoantennas, which are spatially distributed within the metasurface. Therefore, any fabrication imperfections of the nanoantennas could introduce additional noise into the up-converted image (in contrast to conventional schemes). Thus metasurfaces require good fabricated quality and uniformity to obtain clear images from the target in the visible spectrum.
Figure~\ref{fgr:Vis&IRImages}b shows four SFG images generated by our GaAs metasurface and captured on the CMOS camera. An additional long-pass filter was used to filter out the SHG$_{2\omega_p}$ emission. The images correspond to different transverse positions of the target (see Figures~\ref{fgr:Vis&IRImages}b), including the case when the target is fully removed from the path of the signal beam (Figure~\ref{fgr:Vis&IRImages}b-({\em i})). The inhomogeneities observed in Figure~\ref{fgr:Vis&IRImages}b-({\em i}) are due to fabrication defects in the metasurface, as well as inhomogeneities in the pump beam. These inhomogeneities were introduced by the additional 1.6~m distance travelled by the pump beam, with respect to the signal beam (see Supporting Information, Fig. S5), required to temporally synchronize both pulses. Nevertheless, when the target is inserted into the path of the signal beam, the spatial features of the target are preserved in the up-converted images, as shown in Figure~\ref{fgr:Vis&IRImages}b-(\emph{ii-iv}). 

When the pump and signal beam are temporally detuned, the visible images shown in Figure~\ref{fgr:Vis&IRImages}b completely vanished, thus corroborating they are only formed by the SFG process. In our experiment, the resolution of the up-converted images is limited mainly by the SWIR imaging of the target, as seen from the IR image in Figure~\ref{fgr:Vis&IRImages}a. Ultimately, the fundamental limit on the resolution of the visible images is the periodicity and size of the individual nanoantennas. In our case this fundamental resolution is of the order of 750~nm. Overall, the GaAs metasurface enables high-contrast and low-noise IR imaging at room temperature, which are great advantages when compared to other competing technologies.

\section{Conclusions}
In conclusion, we have demonstrated for the first time, to the best of our knowledge, up-conversion of an IR image to visible wavelengths by a resonant ultra-thin GaAs metasurface. The up-conversion is realized by nonlinear wave-mixing of SWIR images with a strong near-IR pump beam within the metasurface. The ultra-fast nonlinear conversion of the sum-frequency process is dramatically enhanced in our 400~nm-thick metasurface due to the excitation of optical resonances at all the three interacting waves. In this way, the IR signal can be easily detected with a simple-uncooled CMOS camera. The realized up-conversion process is parametric and does not exchange energy with the environment, and as such, all spatial information encoded into the IR signal beam is preserved during the up-conversion. Despite different parts of the IR signal beam being up-converted by independent nanoantennas composing the metasurface, the images are well reproduced into the visible, with the ultimate resolution limit being the periodicity of the metasurface.

Unlike current IR cameras, our all-optical approach is not affected by thermal noise and can operate at room temperature using conventional CMOS detectors. Importantly, our metasurface-based IR imaging approach offers novel opportunities, not possible in conventional up-conversion systems where bulky nonlinear crystals are used. For example, the nonlinear wave-mixing can be obtain for counter-propagating pump and signal beams, as well as for incidence at all different angles, as long as the metasurface resonances are excited. Most importantly, multi-color SWIR imaging is also possible by an appropriately designed metasurface. In that case, the designed metasurface would be composed of nanoantennas with different sizes, having resonances at different IR signal wavelengths, while maintaining fixed the resonance of the pump beam. Such metasurface would be able to convert several IR wavelengths to the visible, according to energy conservation observed by the SFG parametric nonlinear process ($\omega_{\rm SFG}$ = $\omega_s$ + $\omega_p$).

We note that our SFG conversion efficiency can be further optimized and enhanced using several strategies, including the use of high-quality factor resonances~\cite{Yang:15,Koshelev:2020:Sci} and materials with higher transparency in the visible region~\cite{Cambiasso:2017:NL}. Additional optimization could also be achieved by employing machine learning approaches for enhancing light-matter interactions~\cite{Xu:2020:AP}. We believe that by enhancing the SFG conversion efficiency, continuous-wave nonlinear up-conversion is within reach.

Our results can directly benefit the development of compact night vision instruments and sensor devices. Notably, the demonstrated  SWIR metasurface imaging devices can be ultra-thin and ultra-compact, be fabricated on flexible substrates, and be fully transparent. In addition, they could offer new functionalities such as multi-color imaging at room-temperature.

\section{METHODS}
\noindent

\textbf{Sample fabrication}.
A (110) GaAs wafer was used as the substrate to epitaxial growth of a 20~nm AlAs sacrificial layer and 400 nm-thick GaAs main layer using Metal-Organic Chemical Vapour Deposition. The AlAs layer was used as a lift-off buffer layer, while the GaAs layer was used to fabricate the nanoantenna arrays. The arrays were defined using electron-beam lithography and sequential inductively coupled plasma etching. Next, the nanoantenna arrays were embedded in a benzocyclobutene (BCB) layer, followed by curing and bonding to a thin glass substrate. Finally, the glass substrate with the GaAs nanoantenna arrays embedded in the BCB layer was lifted off from the GaAs wafer. More details of our fabrication technique can be found in our previous work~\cite{Camacho2016}. The final metasurface consists of a GaAs nanoantenna square array of $30\times30~\mu$m, bonded on a transparent glass substrate. 

\textbf{Numerical simulations}.
We performed numerical calculations using a commercial electromagnetic solver based on a finite element method (COMSOL Multiphysics). The properties of our metasurfaces, consisting of GaAs nanoantennas embedded within a non-dispersive and homogeneous medium (\textit{n}=1.44), are simulated by implementing Floquet boundary conditions to represent an infinite 2D periodic structure. The embedding transparent medium corresponds to the polymer used in our fabrication procedure~\cite{Camacho2016} and the dispersion of the GaAs permittivity was taken from tabulated data~\cite{Aspenes}. For more details see Supporting Information.

\textbf{Experimental setup}.
The optical system used for the nonlinear characterization of the GaAs metasurface is described below. First, the output of a tunable mode-locked Ti:Sapphire laser is split into two beams. One of the beams is coupled to an Optical Parametric Oscillator (OPO) to obtain an IR signal beam, while the other beam is directly used as the pump beam. The tuning range of the pump beam is 740 to 880 nm, while the tuning range of the signal beam is 1000 to 1600 nm. The pump beam passes through an optical delay line (see Supporting Information, Fig. S5), while the signal beam encodes the image of a target through an imaging system (Supporting Information, Fig. S6). The temporal duration of the pump and signal pulses is measured using a frequency-resolved optical grating method (see Supporting Information, Fig. S7a.). The excitation beams are then spatially combined by a dichroic mirror and focused by a lens on the metasurface. The nonlinear emissions generated by the metasurface are collected by an objective lens and sent to a CCD camera or an spectrometer. For more details see Supporting Information.
 
\begin{acknowledgement}

The authors acknowledge the use of the Australian National Fabrication Facility (ANFF), ACT Node. Rocio Camacho-Morales acknowledges a grant from Consejo Nacional de Ciencia y Tecnolog\'{i}a (CONACYT), Mexico; Nikolay Dimitrov and Lyubomir Stoyanov acknowledges a grant from EU Marie-Curie RISE program NOCTURNO; Mohsen Rahmani acknowledges support from the UK Research and Innovation Future Leaders Fellowship (MR/T040513/1). Dragomir N. Neshev acknowledges a grant from the Australian Research Council (CE20010001, DP190101559).

\end{acknowledgement}




\bibliography{references}

\providecommand{\latin}[1]{#1}
\makeatletter
\providecommand{\doi}
  {\begingroup\let\do\@makeother\dospecials
  \catcode`\{=1 \catcode`\}=2 \doi@aux}
\providecommand{\doi@aux}[1]{\endgroup\texttt{#1}}
\makeatother
\providecommand*\mcitethebibliography{\thebibliography}
\csname @ifundefined\endcsname{endmcitethebibliography}
  {\let\endmcitethebibliography\endthebibliography}{}
\begin{mcitethebibliography}{48}
\providecommand*\natexlab[1]{#1}
\providecommand*\mciteSetBstSublistMode[1]{}
\providecommand*\mciteSetBstMaxWidthForm[2]{}
\providecommand*\mciteBstWouldAddEndPuncttrue
  {\def\EndOfBibitem{\unskip.}}
\providecommand*\mciteBstWouldAddEndPunctfalse
  {\let\EndOfBibitem\relax}
\providecommand*\mciteSetBstMidEndSepPunct[3]{}
\providecommand*\mciteSetBstSublistLabelBeginEnd[3]{}
\providecommand*\EndOfBibitem{}
\mciteSetBstSublistMode{f}
\mciteSetBstMaxWidthForm{subitem}{(\alph{mcitesubitemcount})}
\mciteSetBstSublistLabelBeginEnd
  {\mcitemaxwidthsubitemform\space}
  {\relax}
  {\relax}

\bibitem[Schmitt \latin{et~al.}(1998)Schmitt, Xiang, and Yung]{Schmitt:98}
Schmitt,~J.~M.; Xiang,~S.~H.; Yung,~K.~M. Differential absorption imaging with
  optical coherence tomography. \emph{J. Opt. Soc. Am. A} \textbf{1998},
  \emph{15}, 2288--2296\relax
\mciteBstWouldAddEndPuncttrue
\mciteSetBstMidEndSepPunct{\mcitedefaultmidpunct}
{\mcitedefaultendpunct}{\mcitedefaultseppunct}\relax
\EndOfBibitem
\bibitem[Boone \latin{et~al.}(2018)Boone, Zhu, Smith, Todd, and
  Willmott]{BOONE2018601}
Boone,~N.; Zhu,~C.; Smith,~C.; Todd,~I.; Willmott,~J. Thermal near infrared
  monitoring system for electron beam melting with emissivity tracking.
  \emph{Additive Manufacturing} \textbf{2018}, \emph{22}, 601 -- 605\relax
\mciteBstWouldAddEndPuncttrue
\mciteSetBstMidEndSepPunct{\mcitedefaultmidpunct}
{\mcitedefaultendpunct}{\mcitedefaultseppunct}\relax
\EndOfBibitem
\bibitem[Wang and Paliwal(2007)Wang, and Paliwal]{Wang:07}
Wang,~W.; Paliwal,~J. Near-infrared spectroscopy and imaging in food quality
  and safety. \emph{Sensing and Instrumentation for Food Quality and Safety}
  \textbf{2007}, \emph{1}, 193--207\relax
\mciteBstWouldAddEndPuncttrue
\mciteSetBstMidEndSepPunct{\mcitedefaultmidpunct}
{\mcitedefaultendpunct}{\mcitedefaultseppunct}\relax
\EndOfBibitem
\bibitem[K\"{a}llhammer(2006)]{Kallhammer}
K\"{a}llhammer,~J.-E. Imaging: The road ahead for car night-vision.
  \emph{Nature Photonics} \textbf{2006}, 12--13\relax
\mciteBstWouldAddEndPuncttrue
\mciteSetBstMidEndSepPunct{\mcitedefaultmidpunct}
{\mcitedefaultendpunct}{\mcitedefaultseppunct}\relax
\EndOfBibitem
\bibitem[Xia \latin{et~al.}(2015)Xia, Shentu, Shangguan, Xia, Jia, Wang, Zhang,
  Pelc, Fejer, Zhang, Dou, and Pan]{Xia:15}
Xia,~H.; Shentu,~G.; Shangguan,~M.; Xia,~X.; Jia,~X.; Wang,~C.; Zhang,~J.;
  Pelc,~J.~S.; Fejer,~M.~M.; Zhang,~Q.; Dou,~X.; Pan,~J.-W. Long-range
  micro-pulse aerosol lidar at 1.5~$\mu$m with an upconversion single-photon
  detector. \emph{Opt. Lett.} \textbf{2015}, \emph{40}, 1579--1582\relax
\mciteBstWouldAddEndPuncttrue
\mciteSetBstMidEndSepPunct{\mcitedefaultmidpunct}
{\mcitedefaultendpunct}{\mcitedefaultseppunct}\relax
\EndOfBibitem
\bibitem[H{\o}gstedt \latin{et~al.}(2016)H{\o}gstedt, Fix, Wirth, Pedersen, and
  Tidemand-Lichtenberg]{Hogstedt:16}
H{\o}gstedt,~L.; Fix,~A.; Wirth,~M.; Pedersen,~C.; Tidemand-Lichtenberg,~P.
  Upconversion-based lidar measurements of atmospheric {CO}$_2$. \emph{Opt.
  Express} \textbf{2016}, \emph{24}, 5152--5161\relax
\mciteBstWouldAddEndPuncttrue
\mciteSetBstMidEndSepPunct{\mcitedefaultmidpunct}
{\mcitedefaultendpunct}{\mcitedefaultseppunct}\relax
\EndOfBibitem
\bibitem[Midwinter(1968)]{Midwinter}
Midwinter,~J.~E. Image conversion from 1.6 $\mu$ to the visible in lithium
  niobate. \emph{Applied Physics Letters} \textbf{1968}, \emph{12},
  68--70\relax
\mciteBstWouldAddEndPuncttrue
\mciteSetBstMidEndSepPunct{\mcitedefaultmidpunct}
{\mcitedefaultendpunct}{\mcitedefaultseppunct}\relax
\EndOfBibitem
\bibitem[Midwinter(1969)]{Midwinter:69}
Midwinter,~J.~E. Infrared up conversion in Lithium-Niobate with large bandwidth
  and acceptance angle. \emph{Applied Physics Letters} \textbf{1969},
  \emph{14}, 29--32\relax
\mciteBstWouldAddEndPuncttrue
\mciteSetBstMidEndSepPunct{\mcitedefaultmidpunct}
{\mcitedefaultendpunct}{\mcitedefaultseppunct}\relax
\EndOfBibitem
\bibitem[{Andrews}(1970)]{Andrews}
{Andrews},~R. IR image parametric up-conversion. \emph{IEEE Journal of Quantum
  Electronics} \textbf{1970}, \emph{6}, 68--80\relax
\mciteBstWouldAddEndPuncttrue
\mciteSetBstMidEndSepPunct{\mcitedefaultmidpunct}
{\mcitedefaultendpunct}{\mcitedefaultseppunct}\relax
\EndOfBibitem
\bibitem[Weller and Andrews(1970)Weller, and Andrews]{Weller}
Weller,~J.~F.; Andrews,~R.~A. Resolution measurements in parametric
  upconversion of images. \emph{Opto-electronics} \textbf{1970}, \emph{2},
  171--176\relax
\mciteBstWouldAddEndPuncttrue
\mciteSetBstMidEndSepPunct{\mcitedefaultmidpunct}
{\mcitedefaultendpunct}{\mcitedefaultseppunct}\relax
\EndOfBibitem
\bibitem[Voronin \latin{et~al.}(1971)Voronin, Divlekeev, Il'insky, and
  Solomatin]{Voronin}
Voronin,~E.~S.; Divlekeev,~M.~I.; Il'insky,~Y.~A.; Solomatin,~V.~S. The
  influence of the radiation spectrum bandwidth on the resolution of an image
  up-converter. \emph{Opto-electronics} \textbf{1971}, \emph{3}, 153--155\relax
\mciteBstWouldAddEndPuncttrue
\mciteSetBstMidEndSepPunct{\mcitedefaultmidpunct}
{\mcitedefaultendpunct}{\mcitedefaultseppunct}\relax
\EndOfBibitem
\bibitem[Abbas \latin{et~al.}(1976)Abbas, Kostiuk, and Ogilvie]{Abbas:76}
Abbas,~M.~M.; Kostiuk,~T.; Ogilvie,~K.~W. Infrared upconversion for
  astronomical applications. \emph{Appl. Opt.} \textbf{1976}, \emph{15},
  961--970\relax
\mciteBstWouldAddEndPuncttrue
\mciteSetBstMidEndSepPunct{\mcitedefaultmidpunct}
{\mcitedefaultendpunct}{\mcitedefaultseppunct}\relax
\EndOfBibitem
\bibitem[Vasilyev and Kumar(2012)Vasilyev, and Kumar]{Vasilyev:12}
Vasilyev,~M.; Kumar,~P. Frequency up-conversion of quantum images. \emph{Opt.
  Express} \textbf{2012}, \emph{20}, 6644--6656\relax
\mciteBstWouldAddEndPuncttrue
\mciteSetBstMidEndSepPunct{\mcitedefaultmidpunct}
{\mcitedefaultendpunct}{\mcitedefaultseppunct}\relax
\EndOfBibitem
\bibitem[Torregrosa \latin{et~al.}(2015)Torregrosa, Maestre, and
  Capmany]{Torregrosa:15}
Torregrosa,~A.~J.; Maestre,~H.; Capmany,~J. Intra-cavity upconversion to 631 nm
  of images illuminated by an eye-safe {ASE} source at 1550 nm. \emph{Opt.
  Lett.} \textbf{2015}, \emph{40}, 5315--5318\relax
\mciteBstWouldAddEndPuncttrue
\mciteSetBstMidEndSepPunct{\mcitedefaultmidpunct}
{\mcitedefaultendpunct}{\mcitedefaultseppunct}\relax
\EndOfBibitem
\bibitem[Demur \latin{et~al.}(2018)Demur, Garioud, Grisard, Lallier,
  Leviandier, Morvan, Treps, and Fabre]{Demur}
Demur,~R.; Garioud,~R.; Grisard,~A.; Lallier,~E.; Leviandier,~L.; Morvan,~L.;
  Treps,~N.; Fabre,~C. Near-infrared to visible upconversion imaging using a
  broadband pump laser. \emph{Opt. Express} \textbf{2018}, \emph{26},
  13252--13263\relax
\mciteBstWouldAddEndPuncttrue
\mciteSetBstMidEndSepPunct{\mcitedefaultmidpunct}
{\mcitedefaultendpunct}{\mcitedefaultseppunct}\relax
\EndOfBibitem
\bibitem[Vaughan and Trebino(2011)Vaughan, and Trebino]{Vaughan:11}
Vaughan,~P.~M.; Trebino,~R. Optical-parametric-amplification imaging of complex
  objects. \emph{Opt. Express} \textbf{2011}, \emph{19}, 8920--8929\relax
\mciteBstWouldAddEndPuncttrue
\mciteSetBstMidEndSepPunct{\mcitedefaultmidpunct}
{\mcitedefaultendpunct}{\mcitedefaultseppunct}\relax
\EndOfBibitem
\bibitem[Huot \latin{et~al.}(2016)Huot, Moselund, Tidemand-Lichtenberg, Leick,
  and Pedersen]{Huot}
Huot,~L.; Moselund,~P.~M.; Tidemand-Lichtenberg,~P.; Leick,~L.; Pedersen,~C.
  Upconversion imaging using an all-fiber supercontinuum source. \emph{Opt.
  Lett.} \textbf{2016}, \emph{41}, 2466--2469\relax
\mciteBstWouldAddEndPuncttrue
\mciteSetBstMidEndSepPunct{\mcitedefaultmidpunct}
{\mcitedefaultendpunct}{\mcitedefaultseppunct}\relax
\EndOfBibitem
\bibitem[Ashik \latin{et~al.}(2019)Ashik, O'Donnell, Kumar, Ebrahim-Zadeh,
  Tidemand-Lichtenberg, and Pedersen]{Ashik}
Ashik,~A.~S.; O'Donnell,~C.~F.; Kumar,~S.~C.; Ebrahim-Zadeh,~M.;
  Tidemand-Lichtenberg,~P.; Pedersen,~C. Mid-infrared upconversion imaging
  using femtosecond pulses. \emph{Photon. Res.} \textbf{2019}, \emph{7},
  783--791\relax
\mciteBstWouldAddEndPuncttrue
\mciteSetBstMidEndSepPunct{\mcitedefaultmidpunct}
{\mcitedefaultendpunct}{\mcitedefaultseppunct}\relax
\EndOfBibitem
\bibitem[Jacobo \latin{et~al.}(2005)Jacobo, Colet, Scotto, and
  San~Miguel]{Jacobo}
Jacobo,~A.; Colet,~P.; Scotto,~P.; San~Miguel,~M. Use of nonlinear properties
  of intracavity type {II} second harmonic generation for image processing.
  \emph{Applied Physics B} \textbf{2005}, \emph{81}, 955--962\relax
\mciteBstWouldAddEndPuncttrue
\mciteSetBstMidEndSepPunct{\mcitedefaultmidpunct}
{\mcitedefaultendpunct}{\mcitedefaultseppunct}\relax
\EndOfBibitem
\bibitem[Neshev and Aharonovich(2018)Neshev, and Aharonovich]{Neshev:18}
Neshev,~D.; Aharonovich,~I. Optical metasurfaces: new generation building
  blocks for multi-functional optics. \emph{Light: Science \& Applications}
  \textbf{2018}, \emph{7}, 58\relax
\mciteBstWouldAddEndPuncttrue
\mciteSetBstMidEndSepPunct{\mcitedefaultmidpunct}
{\mcitedefaultendpunct}{\mcitedefaultseppunct}\relax
\EndOfBibitem
\bibitem[Rahmani \latin{et~al.}(2018)Rahmani, Leo, Brener, Zayats, Maier,
  Angelis, Tan, Gili, Karouta, Oulton, Vora, Lysevych, Staude, Xu,
  Miroshnichenko, Jagadish, and Neshev]{Rahmani:18}
Rahmani,~M. \latin{et~al.}  Nonlinear frequency conversion in optical
  nanoantennas and metasurfaces: materials evolution and fabrication.
  \emph{Opto-Electronic Advances} \textbf{2018}, \emph{1}, 180021\relax
\mciteBstWouldAddEndPuncttrue
\mciteSetBstMidEndSepPunct{\mcitedefaultmidpunct}
{\mcitedefaultendpunct}{\mcitedefaultseppunct}\relax
\EndOfBibitem
\bibitem[De~Angelis \latin{et~al.}(2020)De~Angelis, Leo, and
  Neshev]{deAngelis:2020nonlinear}
De~Angelis,~C.; Leo,~G.; Neshev,~D. \emph{Nonlinear Meta-Optics}; CRC Press,
  2020\relax
\mciteBstWouldAddEndPuncttrue
\mciteSetBstMidEndSepPunct{\mcitedefaultmidpunct}
{\mcitedefaultendpunct}{\mcitedefaultseppunct}\relax
\EndOfBibitem
\bibitem[Shcherbakov \latin{et~al.}(2014)Shcherbakov, Neshev, Hopkins,
  Shorokhov, Staude, Melik-Gaykazyan, Decker, Ezhov, Miroshnichenko, Brener,
  Fedyanin, and Kivshar]{Shcherbakov:2014}
Shcherbakov,~M.~R.; Neshev,~D.~N.; Hopkins,~B.; Shorokhov,~A.~S.; Staude,~I.;
  Melik-Gaykazyan,~E.~V.; Decker,~M.; Ezhov,~A.~A.; Miroshnichenko,~A.~E.;
  Brener,~I.; Fedyanin,~A.~A.; Kivshar,~Y.~S. Enhanced Third-Harmonic
  Generation in Silicon Nanoparticles Driven by Magnetic Response. \emph{Nano
  Lett.} \textbf{2014}, \emph{14}, 6488--6492\relax
\mciteBstWouldAddEndPuncttrue
\mciteSetBstMidEndSepPunct{\mcitedefaultmidpunct}
{\mcitedefaultendpunct}{\mcitedefaultseppunct}\relax
\EndOfBibitem
\bibitem[Yang \latin{et~al.}(2015)Yang, Wang, Boulesbaa, Kravchenko, Briggs,
  Puretzky, Geohegan, and Valentine]{Yang:15}
Yang,~Y.; Wang,~W.; Boulesbaa,~A.; Kravchenko,~I.~I.; Briggs,~D.~P.;
  Puretzky,~A.; Geohegan,~D.; Valentine,~J. Nonlinear Fano-Resonant Dielectric
  Metasurfaces. \emph{Nano Lett.} \textbf{2015}, \emph{15}, 7388--7393\relax
\mciteBstWouldAddEndPuncttrue
\mciteSetBstMidEndSepPunct{\mcitedefaultmidpunct}
{\mcitedefaultendpunct}{\mcitedefaultseppunct}\relax
\EndOfBibitem
\bibitem[Tong \latin{et~al.}(2016)Tong, Gong, Liu, Yuan, Huang, Xia, and
  Wang]{Tong:16}
Tong,~W.; Gong,~C.; Liu,~X.; Yuan,~S.; Huang,~Q.; Xia,~J.; Wang,~Y. Enhanced
  third harmonic generation in a silicon metasurface using trapped mode.
  \emph{Opt. Express} \textbf{2016}, \emph{24}, 19661--19670\relax
\mciteBstWouldAddEndPuncttrue
\mciteSetBstMidEndSepPunct{\mcitedefaultmidpunct}
{\mcitedefaultendpunct}{\mcitedefaultseppunct}\relax
\EndOfBibitem
\bibitem[Semmlinger \latin{et~al.}(2019)Semmlinger, Zhang, Tseng, Huang, Yang,
  Tsai, Nordlander, and Halas]{Semmlinger:19}
Semmlinger,~M.; Zhang,~M.; Tseng,~M.~L.; Huang,~T.-T.; Yang,~J.; Tsai,~D.~P.;
  Nordlander,~P.; Halas,~N.~J. Generating Third Harmonic Vacuum Ultraviolet
  Light with a {TiO}$_2$ Metasurface. \emph{Nano Letters} \textbf{2019},
  \emph{19}, 8972--8978\relax
\mciteBstWouldAddEndPuncttrue
\mciteSetBstMidEndSepPunct{\mcitedefaultmidpunct}
{\mcitedefaultendpunct}{\mcitedefaultseppunct}\relax
\EndOfBibitem
\bibitem[Ohashi \latin{et~al.}(1993)Ohashi, Kondo, Ito, Fukatsu, Shiraki,
  Kumata, and Kano]{Ohashi:1993}
Ohashi,~M.; Kondo,~T.; Ito,~R.; Fukatsu,~S.; Shiraki,~Y.; Kumata,~K.;
  Kano,~S.~S. Determination of quadratic nonlinear optical coefficient of
  {Al$_x$Ga$_{1-x}$As} system by the method of reflected second harmonics.
  \emph{J. Appl. Phys.} \textbf{1993}, \emph{74}, 596--601\relax
\mciteBstWouldAddEndPuncttrue
\mciteSetBstMidEndSepPunct{\mcitedefaultmidpunct}
{\mcitedefaultendpunct}{\mcitedefaultseppunct}\relax
\EndOfBibitem
\bibitem[Gili \latin{et~al.}(2016)Gili, Carletti, Locatelli, Rocco, Finazzi,
  Ghirardini, Favero, Gomez, Lemaître, Celebrano, De~Angelis, and
  Leo]{Gili:2016:OE}
Gili,~V.~F.; Carletti,~L.; Locatelli,~A.; Rocco,~D.; Finazzi,~M.;
  Ghirardini,~L.; Favero,~I.; Gomez,~C.; Lemaître,~A.; Celebrano,~M.;
  De~Angelis,~C.; Leo,~G. Monolithic AlGaAs second-harmonic nanoantennas.
  \emph{Opt. Express} \textbf{2016}, \emph{24}, 15965--15971\relax
\mciteBstWouldAddEndPuncttrue
\mciteSetBstMidEndSepPunct{\mcitedefaultmidpunct}
{\mcitedefaultendpunct}{\mcitedefaultseppunct}\relax
\EndOfBibitem
\bibitem[Liu \latin{et~al.}(2016)Liu, Sinclair, Saravi, Keeler, Yang, Reno,
  Peake, Setzpfandt, Staude, Pertsch, and Brener]{Liu:2016:NL}
Liu,~S.; Sinclair,~M.~B.; Saravi,~S.; Keeler,~G.~A.; Yang,~Y.; Reno,~J.;
  Peake,~G.~M.; Setzpfandt,~F.; Staude,~I.; Pertsch,~T.; Brener,~I. Resonantly
  Enhanced Second-Harmonic Generation Using III–V Semiconductor
  All-Dielectric Metasurfaces. \emph{Nano Lett.} \textbf{2016}, \emph{16},
  5426--5432\relax
\mciteBstWouldAddEndPuncttrue
\mciteSetBstMidEndSepPunct{\mcitedefaultmidpunct}
{\mcitedefaultendpunct}{\mcitedefaultseppunct}\relax
\EndOfBibitem
\bibitem[Camacho-Morales \latin{et~al.}(2016)Camacho-Morales, Rahmani, Kruk,
  Wang, Xu, Smirnova, Solntsev, Miroshnichenko, Tan, Karouta, Naureen, Vora,
  Carletti, De~Angelis, Jagadish, Kivshar, and Neshev]{Camacho2016}
Camacho-Morales,~R. \latin{et~al.}  Nonlinear Generation of Vector Beams From
  AlGaAs Nanoantennas. \emph{Nano Letters} \textbf{2016}, \emph{16},
  7191--7197\relax
\mciteBstWouldAddEndPuncttrue
\mciteSetBstMidEndSepPunct{\mcitedefaultmidpunct}
{\mcitedefaultendpunct}{\mcitedefaultseppunct}\relax
\EndOfBibitem
\bibitem[Liu \latin{et~al.}(2018)Liu, Vabishchevich, Vaskin, Reno, Keeler,
  Sinclair, Staude, and Brener]{Liu:2018:NATCOMM}
Liu,~S.; Vabishchevich,~P.~P.; Vaskin,~A.; Reno,~J.~L.; Keeler,~G.~A.;
  Sinclair,~M.~B.; Staude,~I.; Brener,~I. An all-dielectric metasurface as a
  broadband optical frequency mixer. \emph{Nature Communications}
  \textbf{2018}, \emph{9}, 2507\relax
\mciteBstWouldAddEndPuncttrue
\mciteSetBstMidEndSepPunct{\mcitedefaultmidpunct}
{\mcitedefaultendpunct}{\mcitedefaultseppunct}\relax
\EndOfBibitem
\bibitem[Ha \latin{et~al.}(2018)Ha, Fu, Emani, Pan, Bakker,
  Paniagua-Domínguez, and Kuznetsov]{Ha:18}
Ha,~S.~T.; Fu,~Y.~H.; Emani,~N.~K.; Pan,~Z.; Bakker,~R.~M.;
  Paniagua-Domínguez,~R.; Kuznetsov,~A.~I. Directional lasing in resonant
  semiconductor nanoantenna arrays. \emph{Nature Nanotechnology} \textbf{2018},
  \emph{13}, 1042--1047\relax
\mciteBstWouldAddEndPuncttrue
\mciteSetBstMidEndSepPunct{\mcitedefaultmidpunct}
{\mcitedefaultendpunct}{\mcitedefaultseppunct}\relax
\EndOfBibitem
\bibitem[L\"ochner \latin{et~al.}(2018)L\"ochner, Fedotova, Liu, Keeler, Peake,
  Saravi, Shcherbakov, Burger, Fedyanin, Brener, Pertsch, Setzpfandt, and
  Staude]{Lochner:2018:ACSPh}
L\"ochner,~F. J.~F.; Fedotova,~A.~N.; Liu,~S.; Keeler,~G.~A.; Peake,~G.~M.;
  Saravi,~S.; Shcherbakov,~M.~R.; Burger,~S.; Fedyanin,~A.~A.; Brener,~I.;
  Pertsch,~T.; Setzpfandt,~F.; Staude,~I. Polarization-Dependent Second
  Harmonic Diffraction from Resonant GaAs Metasurfaces. \emph{ACS Photon.}
  \textbf{2018}, \emph{5}, 1786--1793\relax
\mciteBstWouldAddEndPuncttrue
\mciteSetBstMidEndSepPunct{\mcitedefaultmidpunct}
{\mcitedefaultendpunct}{\mcitedefaultseppunct}\relax
\EndOfBibitem
\bibitem[Vabishchevich \latin{et~al.}(2018)Vabishchevich, Liu, Sinclair,
  Keeler, Peake, and Brener]{Vabishchevich}
Vabishchevich,~P.~P.; Liu,~S.; Sinclair,~M.~B.; Keeler,~G.~A.; Peake,~G.~M.;
  Brener,~I. Enhanced Second-Harmonic Generation Using Broken Symmetry III–V
  Semiconductor Fano Metasurfaces. \emph{ACS Photonics} \textbf{2018},
  \emph{5}, 1685--1690\relax
\mciteBstWouldAddEndPuncttrue
\mciteSetBstMidEndSepPunct{\mcitedefaultmidpunct}
{\mcitedefaultendpunct}{\mcitedefaultseppunct}\relax
\EndOfBibitem
\bibitem[Marino \latin{et~al.}(2019)Marino, Gigli, Rocco, Lemaître, Favero,
  De~Angelis, and Leo]{Marino:2019:ACSPh}
Marino,~G.; Gigli,~C.; Rocco,~D.; Lemaître,~A.; Favero,~I.; De~Angelis,~C.;
  Leo,~G. Zero-Order Second Harmonic Generation from AlGaAs-on-Insulator
  Metasurfaces. \emph{ACS Photon.} \textbf{2019}, \emph{6}, 1226--1231\relax
\mciteBstWouldAddEndPuncttrue
\mciteSetBstMidEndSepPunct{\mcitedefaultmidpunct}
{\mcitedefaultendpunct}{\mcitedefaultseppunct}\relax
\EndOfBibitem
\bibitem[Rocco \latin{et~al.}(2020)Rocco, Gigli, Carletti, Marino, Vincenti,
  Leo, and Angelis]{Rocco:2020:IEEE}
Rocco,~D.; Gigli,~C.; Carletti,~L.; Marino,~G.; Vincenti,~M.~A.; Leo,~G.;
  Angelis,~C.~D. Vertical Second Harmonic Generation in Asymmetric Dielectric
  Nanoantennas. \emph{IEEE Photonics Journal} \textbf{2020}, \emph{12},
  1--7\relax
\mciteBstWouldAddEndPuncttrue
\mciteSetBstMidEndSepPunct{\mcitedefaultmidpunct}
{\mcitedefaultendpunct}{\mcitedefaultseppunct}\relax
\EndOfBibitem
\bibitem[Sautter \latin{et~al.}(2019)Sautter, Xu, Miroshnichenko, Lysevych,
  Volkovskaya, Smirnova, Camacho-Morales, Zangeneh~Kamali, Karouta, Vora, Tan,
  Kauranen, Staude, Jagadish, Neshev, and Rahmani]{Sautter:2019:NL}
Sautter,~J.~D. \latin{et~al.}  Tailoring Second-Harmonic Emission from
  (111)-{G}a{A}s Nanoantennas. \emph{Nano Lett.} \textbf{2019}, \emph{19},
  3905--3911\relax
\mciteBstWouldAddEndPuncttrue
\mciteSetBstMidEndSepPunct{\mcitedefaultmidpunct}
{\mcitedefaultendpunct}{\mcitedefaultseppunct}\relax
\EndOfBibitem
\bibitem[Xu \latin{et~al.}(2020)Xu, Saerens, Timofeeva, Smirnova, Volkovskaya,
  Lysevych, Camacho-Morales, Cai, Zangeneh~Kamali, Huang, Karouta, Tan,
  Jagadish, Miroshnichenko, Grange, Neshev, and Rahmani]{Xu:20}
Xu,~L. \latin{et~al.}  Forward and Backward Switching of Nonlinear
  Unidirectional Emission from GaAs Nanoantennas. \emph{ACS Nano}
  \textbf{2020}, \emph{14}, 1379--1389\relax
\mciteBstWouldAddEndPuncttrue
\mciteSetBstMidEndSepPunct{\mcitedefaultmidpunct}
{\mcitedefaultendpunct}{\mcitedefaultseppunct}\relax
\EndOfBibitem
\bibitem[Krieg and Adomeit(2019)Krieg, and Adomeit]{Krieg:19}
Krieg,~J.; Adomeit,~U. Comparative long-time visible and shortwave infrared
  night illumination measurements. \emph{Appl. Opt.} \textbf{2019}, \emph{58},
  9876--9882\relax
\mciteBstWouldAddEndPuncttrue
\mciteSetBstMidEndSepPunct{\mcitedefaultmidpunct}
{\mcitedefaultendpunct}{\mcitedefaultseppunct}\relax
\EndOfBibitem
\bibitem[Celebrano \latin{et~al.}(2015)Celebrano, Wu, Baselli, Großmann,
  Biagioni, Locatelli, De~Angelis, Cerullo, Osellame, Hecht, Duò, Ciccacci,
  and Finazzi]{Celebrano}
Celebrano,~M.; Wu,~X.; Baselli,~M.; Großmann,~S.; Biagioni,~P.; Locatelli,~A.;
  De~Angelis,~C.; Cerullo,~G.; Osellame,~R.; Hecht,~B.; Duò,~L.; Ciccacci,~F.;
  Finazzi,~M. Mode matching in multiresonant plasmonic nanoantennas for
  enhanced second harmonic generation. \emph{Nature Nanotechnology}
  \textbf{2015}, \emph{10}, 412--417\relax
\mciteBstWouldAddEndPuncttrue
\mciteSetBstMidEndSepPunct{\mcitedefaultmidpunct}
{\mcitedefaultendpunct}{\mcitedefaultseppunct}\relax
\EndOfBibitem
\bibitem[Colom \latin{et~al.}(2019)Colom, Xu, Marini, Bedu, Ozerov, Begou,
  Lumeau, Miroshnishenko, Neshev, Kuhlmey, Palomba, and Bonod]{Colom}
Colom,~R.; Xu,~L.; Marini,~L.; Bedu,~F.; Ozerov,~I.; Begou,~T.; Lumeau,~J.;
  Miroshnishenko,~A.~E.; Neshev,~D.; Kuhlmey,~B.~T.; Palomba,~S.; Bonod,~N.
  Enhanced Four-Wave Mixing in Doubly Resonant Si Nanoresonators. \emph{ACS
  Photonics} \textbf{2019}, \emph{6}, 1295--1301\relax
\mciteBstWouldAddEndPuncttrue
\mciteSetBstMidEndSepPunct{\mcitedefaultmidpunct}
{\mcitedefaultendpunct}{\mcitedefaultseppunct}\relax
\EndOfBibitem
\bibitem[Harutyunyan \latin{et~al.}(2012)Harutyunyan, Volpe, Quidant, and
  Novotny]{Harutyunyan}
Harutyunyan,~H.; Volpe,~G.; Quidant,~R.; Novotny,~L. Enhancing the Nonlinear
  Optical Response Using Multifrequency Gold-Nanowire Antennas. \emph{Phys.
  Rev. Lett.} \textbf{2012}, \emph{108}, 217403\relax
\mciteBstWouldAddEndPuncttrue
\mciteSetBstMidEndSepPunct{\mcitedefaultmidpunct}
{\mcitedefaultendpunct}{\mcitedefaultseppunct}\relax
\EndOfBibitem
\bibitem[Antonucci \latin{et~al.}(2012)Antonucci, Solinas, Bonvalet, and
  Joffre]{Antonucci:12}
Antonucci,~L.; Solinas,~X.; Bonvalet,~A.; Joffre,~M. Asynchronous optical
  sampling with arbitrary detuning between laser repetition rates. \emph{Opt.
  Express} \textbf{2012}, \emph{20}, 17928--17937\relax
\mciteBstWouldAddEndPuncttrue
\mciteSetBstMidEndSepPunct{\mcitedefaultmidpunct}
{\mcitedefaultendpunct}{\mcitedefaultseppunct}\relax
\EndOfBibitem
\bibitem[Koshelev \latin{et~al.}(2020)Koshelev, Kruk, Melik-Gaykazyan, Choi,
  Bogdanov, Park, and Kivshar]{Koshelev:2020:Sci}
Koshelev,~K.; Kruk,~S.; Melik-Gaykazyan,~E.; Choi,~J.-H.; Bogdanov,~A.;
  Park,~H.-G.; Kivshar,~Y. Subwavelength dielectric resonators for nonlinear
  nanophotonics. \emph{Science} \textbf{2020}, \emph{367}, 288--292\relax
\mciteBstWouldAddEndPuncttrue
\mciteSetBstMidEndSepPunct{\mcitedefaultmidpunct}
{\mcitedefaultendpunct}{\mcitedefaultseppunct}\relax
\EndOfBibitem
\bibitem[Cambiasso \latin{et~al.}(2017)Cambiasso, Grinblat, Li, Rakovich,
  Cortés, and Maier]{Cambiasso:2017:NL}
Cambiasso,~J.; Grinblat,~G.; Li,~Y.; Rakovich,~A.; Cortés,~E.; Maier,~S.~A.
  Bridging the Gap between Dielectric Nanophotonics and the Visible Regime with
  Effectively Lossless Gallium Phosphide Antennas. \emph{Nano Letters}
  \textbf{2017}, \emph{17}, 1219--1225\relax
\mciteBstWouldAddEndPuncttrue
\mciteSetBstMidEndSepPunct{\mcitedefaultmidpunct}
{\mcitedefaultendpunct}{\mcitedefaultseppunct}\relax
\EndOfBibitem
\bibitem[Xu \latin{et~al.}(2020)Xu, Rahmani, Ma, Smirnova, Kamali, Deng,
  Chiang, Huang, Zhang, Gould, Neshev, and Miroshnichenko]{Xu:2020:AP}
Xu,~L.; Rahmani,~M.; Ma,~Y.; Smirnova,~D.~A.; Kamali,~K.~Z.; Deng,~F.;
  Chiang,~Y.~K.; Huang,~L.; Zhang,~H.; Gould,~S.; Neshev,~D.~N.;
  Miroshnichenko,~A.~E. {Enhanced light–matter interactions in dielectric
  nanostructures via machine-learning approach}. \emph{Advanced Photonics}
  \textbf{2020}, \emph{2}, 1 -- 11\relax
\mciteBstWouldAddEndPuncttrue
\mciteSetBstMidEndSepPunct{\mcitedefaultmidpunct}
{\mcitedefaultendpunct}{\mcitedefaultseppunct}\relax
\EndOfBibitem
\bibitem[Aspnes \latin{et~al.}(1986)Aspnes, Kelso, Logan, and Bhat]{Aspenes}
Aspnes,~D.~E.; Kelso,~S.~M.; Logan,~R.~A.; Bhat,~R. Optical properties of
  {Al$_x$Ga $_{1-x}$As}. \emph{Journal of Applied Physics} \textbf{1986},
  \emph{60}, 754--767\relax
\mciteBstWouldAddEndPuncttrue
\mciteSetBstMidEndSepPunct{\mcitedefaultmidpunct}
{\mcitedefaultendpunct}{\mcitedefaultseppunct}\relax
\EndOfBibitem
\end{mcitethebibliography}

\beginsupplement 

\section{Supporting Information}

\subsection{Numerical calculations}
The linear and nonlinear optical response of the (110) GaAs metasurface is numerically modeled by using the Finite Element Method in Comsol Multiphysics. The nonlinear response is obtained in a two-step approach. First, we calculate the nonlinear polarization response of the metasurface \textbf{P}($\omega_3$) resulting from the incident pump and signal frequency beams. Then, we employ the nonlinear polarization as the source to calculate the SFG, through the induced nonlinear current \textbf{J}($\omega_3$). We define the \textit{i}-th component of the nonlinear electric polarization vector at the angular frequency $\omega_3$ as
\begin{equation}
P_i(\omega_3) = \epsilon_0 \chi^{(2)} [\textit{E}_j(\omega_1)\textit{E}_k(\omega_2) + \textit{E}_k(\omega_1)\textit{E}_j(\omega_2)],    
\end{equation}
with \textit{i} $\ne$ \textit{j} $\ne$ \textit{i} due to the zinc-blende crystal structure of GaAs. Here $\epsilon_0$ is the vacuum permittivity, \textit{E}$_j$($\omega_1$) is the \textit{j}-th component of the electric field at the angular frequency $\omega_1$ and \textit{E}$_k$($\omega_2$) is the \textit{k}-th component of the electric field at the angular frequency $\omega_2$. The angular frequencies $\omega_1$, $\omega_2$ and $\omega_3$ correspond to the wavelength of the signal at 1530 nm, the pump at 860 nm, and the SFG at 550 nm, respectively. The GaAs metasurface is simulated by implementing Floquet boundary conditions to mimic an infinite 2D periodic structure.

\begin{figure}[htbp]
\centering
\includegraphics[width=0.95\linewidth]{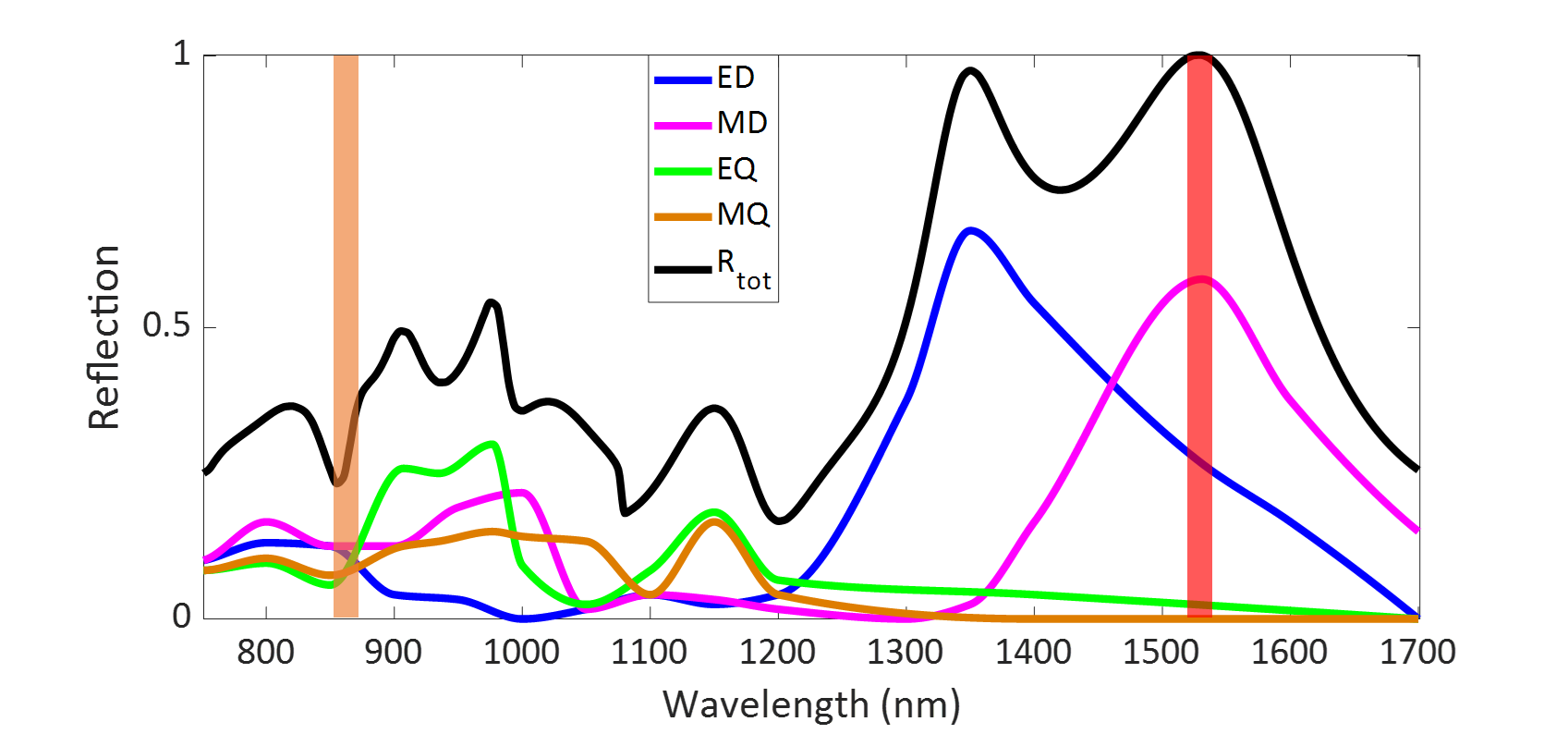}
\caption{Calculated total reflection spectrum (black solid line, R$_{tot}$) and multipolar decomposition of GaAs metasurface, as a function of incident wavelength. The multipolar decomposition shows the main contributions of the reflection spectrum, namely electric dipole (ED), magnetic dipole (MD), electric quadrupole (EQ) and magnetic quadruple (MQ). The orange and red vertical lines indicate the wavelength of the pump (860~nm) and signal (1530~nm) beams, respectively.}
\label{fig:multipolar}
\end{figure}

Figure~\ref{fig:multipolar} shows the calculated reflection of our GaAs metasurface ($P=750$~nm, $r+225$~nm and $h=400$~nm), as a function of incident wavelength. As shown in Figure~\ref{fig:multipolar}, at long wavelengths the reflection spectrum has mainly electric and magnetic dipole contributions, while for shorter wavelengths the reflection spectrum has additional contributions from electric and magnetic quadrupole modes. Thus, at the incident IR signal beam ($\lambda$=1530~nm) the main contributions are electric and magnetic dipole modes. On the other hand, at the incident pump beam ($\lambda$=860~nm), the resonant behavior of the metasurface is explained by the combination of dipoles and quadrupole modes of electric and magnetic nature.

\begin{figure}[htbp]
\centering
\includegraphics[width=0.85\linewidth]{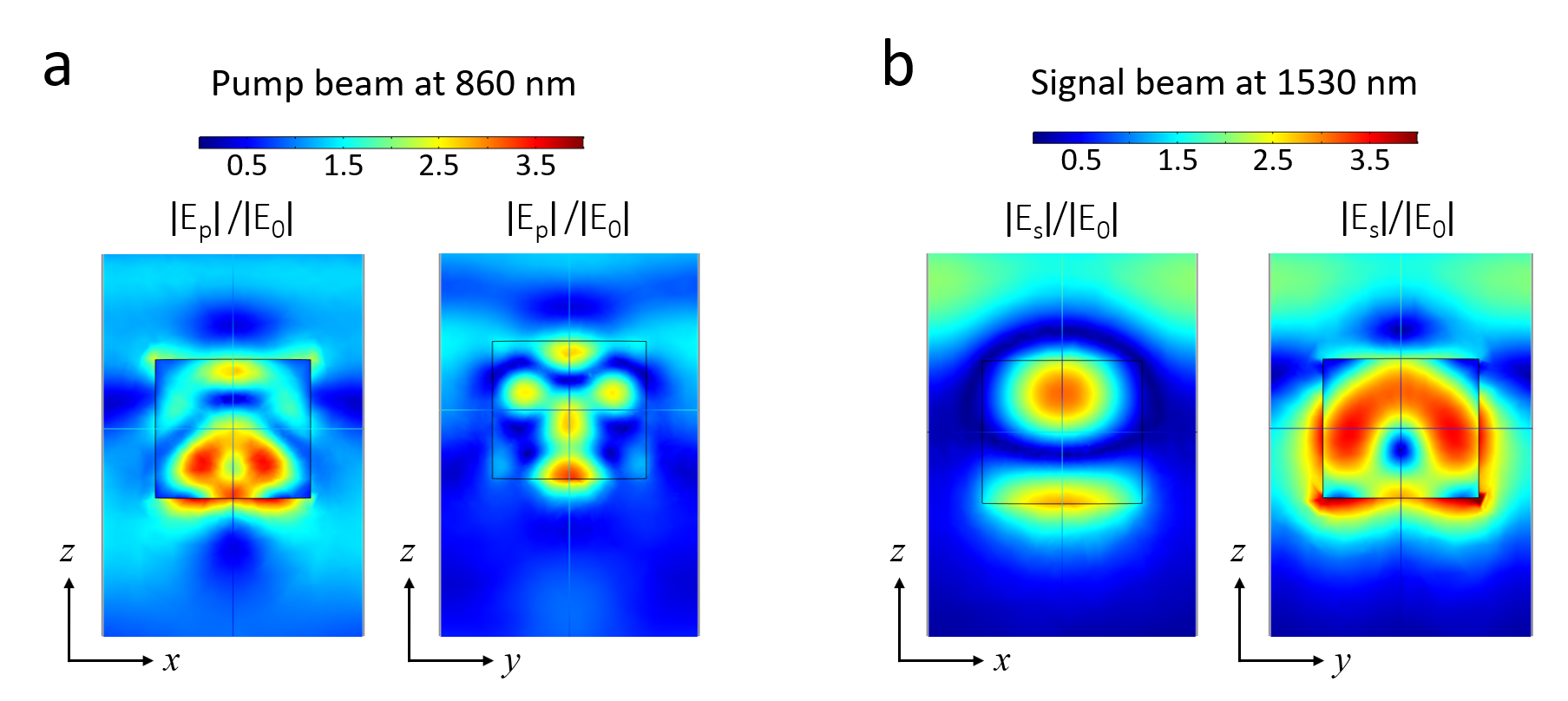}
\caption{Calculated modulus of electric field distribution through the center of GaAs metasurface unitary cell  (\textit{r}=225~nm, \textit{h}=400~nm) at an incident wavelength of (a) 860 and (b) 1530~nm.}
\label{fig:pump and signal profiles}
\end{figure}

\begin{figure}[htbp]
\centering
\includegraphics[width=0.85\linewidth]{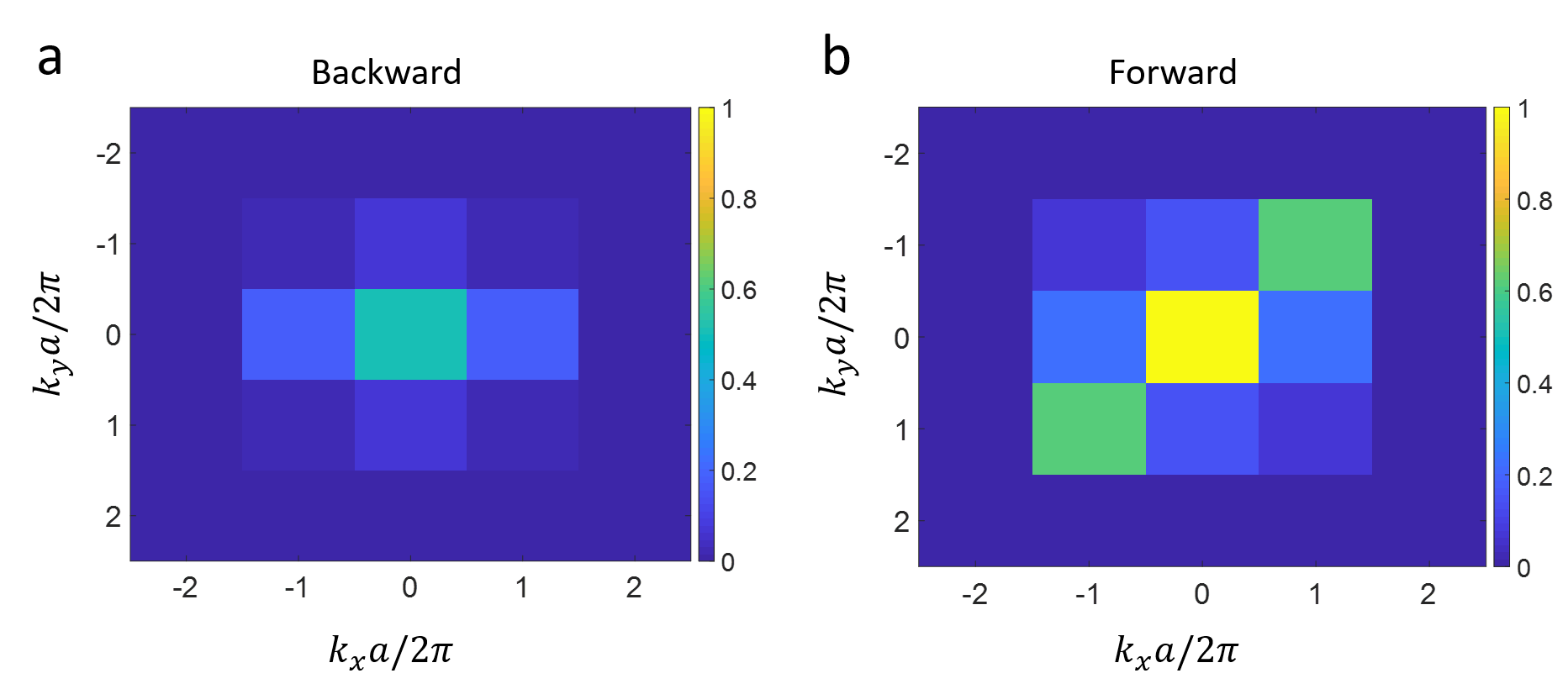}
\caption{Far-field diffraction distribution of SFG calculated in the (a) backward and (b) forward direction to the GaAs metasurface, when simultaneously excited by a pump beam at 860~nm and a signal beam at 1530~nm. The color scale of SFG intensity is shown at the right side of each plot.}
\label{fig:SFG-diffraction-orders}
\end{figure}

When considering the lattice effects of the GaAs metasurface, the sum-frequency emission will be shaped into different diffraction orders, depending on the periodicity \textit{P} of the metasurface. Figure~\ref{fig:SFG-diffraction-orders} shows the sum-frequency diffraction coefficients of our GaAs metasurface generated by the simultaneous incidence of an IR signal beam at 1530 nm and a pump at 860 nm. The SFG diffraction coefficients are calculated by performing the Fourier transform of the sum-frequency near field in both directions, backward and forward. As indicated by the color intensity scale in Figure~\ref{fig:SFG-diffraction-orders}a and b, in each direction the strongest SFG emission corresponds to the zero-th diffraction order. In both directions, the first diffraction orders in the \textit{x}- and \textit{y}-directions have lower intensity than the zero-th diffraction order. Overall, the forward SFG intensity is stronger than the backward SFG. In our metasurface, there are no second diffraction orders for the sum-frequency emission. 

\begin{figure}[htbp]
\centering
\includegraphics[width=0.65\linewidth]{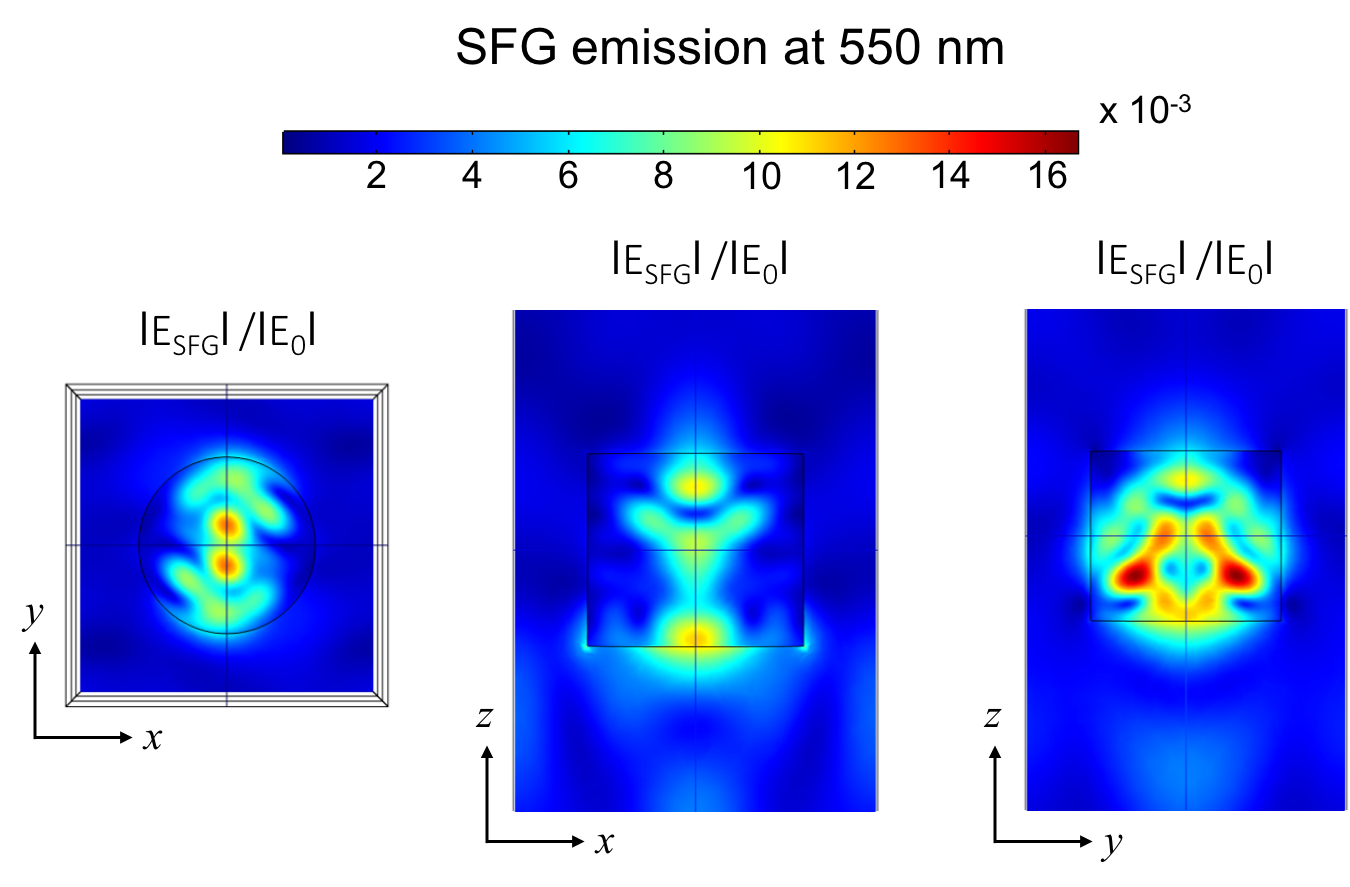}
\caption{Calculated modulus of the sum-frequency electric field distribution, across the center of the \textit{xy}-, \textit{xz}- and \textit{yz}-plane of GaAs metasurface unitary cell ($r=225$~nm, $h=400$~nm). The SFG field is generated at 550 nm by the simultaneous incidence of pump and signal beams at 860 and 1530~nm, respectively. Here \textit{E}$_0$ refers to the amplitude of the incident pump beam.}
\label{fig:SFG profile}
\end{figure}

\subsection{Experimental setup and measurements}

\begin{figure}[htbp]
\centering
\includegraphics[width=0.9\linewidth]{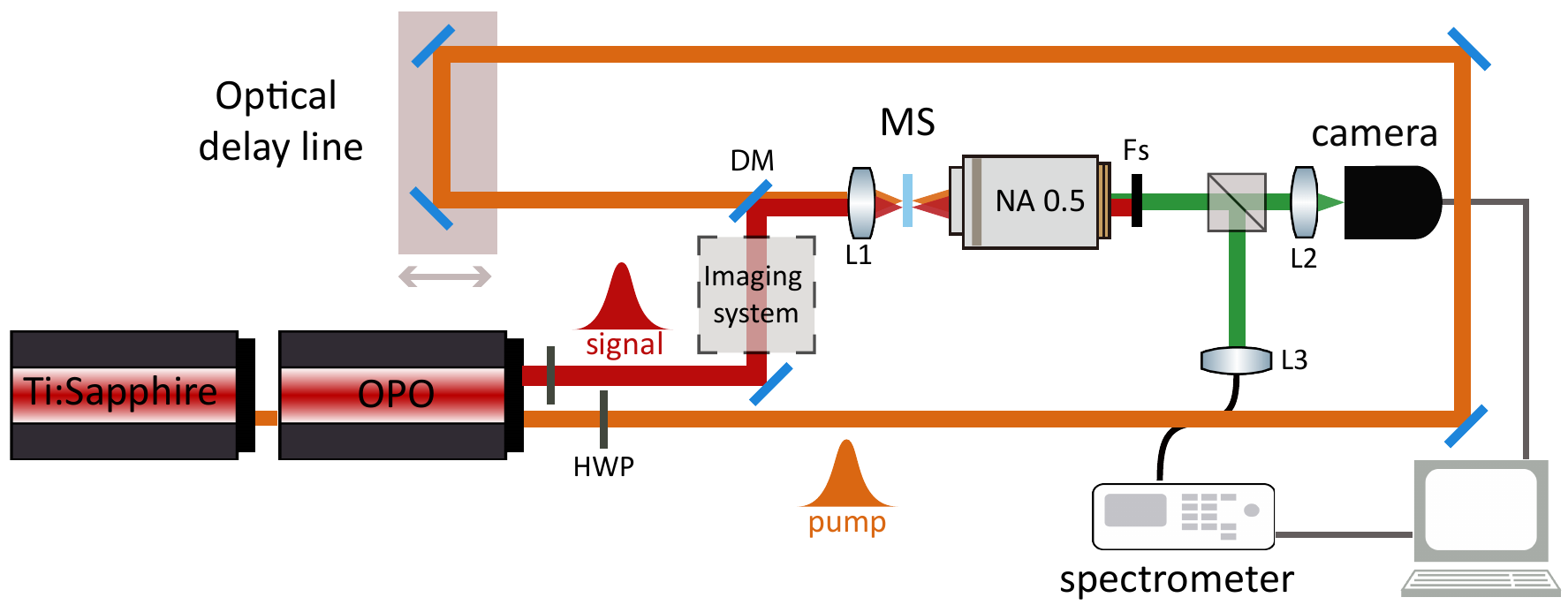}
\caption{Schematic of optical system used to study the nonlinear emissions generated by (110) GaAs metasurface. The schematic shows the optical path of the signal and pump pulsed laser beams employed to excite the metasurface and generate sum-frequency emission at green wavelengths. The IR up-conversion imaging is performed by thusing the imaging system in the IR signal beam. In the schematic, the focused pump and signal beams are not spatially overlapped only for visualization purposes.}
\label{fig:setup}
\end{figure}

The schematic of the optical system used to characterize the nonlinear optical response of the GaAs metasurface is shown in Figure~\ref{fig:setup}. A Ti:Sapphire laser (Coherent, Chameleon Ultra II) pumps an optical parametric oscillator cavity (Coherent, Chameleon Compact OPO), which provides a short-wave IR signal beam and an unconverted beam with the same wavelength as the Ti:Sapphire pump. Therefore, at the output of the OPO cavity two beams with the same polarization (horizontal linear polarization) and repetition rate (80 MHz), but different wavelength, are used as the signal and pump excitation beams. The polarization angle of the signal and pump beams can be controlled separately by two half-wave plates, \textit{HWP} (Thorlabs, AHWP10M-980 and AHWP05M-1600). Temporal synchronization of the signal and pump pulsed laser beams is achieved by tuning an optical delay line which consists of two mirrors sitting at about 90 degrees to each other, mounted in a travel translation stage (Thorlabs, PT1/M). The signal and pump beams are spatially combined by a dichroic mirror, \textit{DM} (Thorlabs, Low-GDD Ultrafast Mirror) and then focused at the same position on the GaAs metasurface, \textit{MS} by a plano-convex lens, \textit{L1} with \textit{f}=50 mm. The nonlinear radiation generated by the GaAs metasurface, together with the transmitted excitation beams, are collected by a 100X objective lens with a NA of 0.5 (Mitutoyo, Plan Apo NIR Infinity). At the back aperture of the objective lens, two short-pass filters, \textit{Fs} with cut-off wavelengths of 600 and 800 nm (Thorlabs, FESH0600 and FESH0800) are used to filter-out the transmitted excitation beams. The transmitted nonlinear emissions are either focused by a plano-convex lens, \textit{L2} with \textit{f}=75 mm in the CCD of a camera (Thorlabs, DCC1545M) or coupled by a fiber collimator lens, \textit{L3} to an optical fiber connected to a visible spectrometer (Ocean Optics, QE65000). 

\begin{figure}[htbp]
\centering
\includegraphics[width=0.6\linewidth]{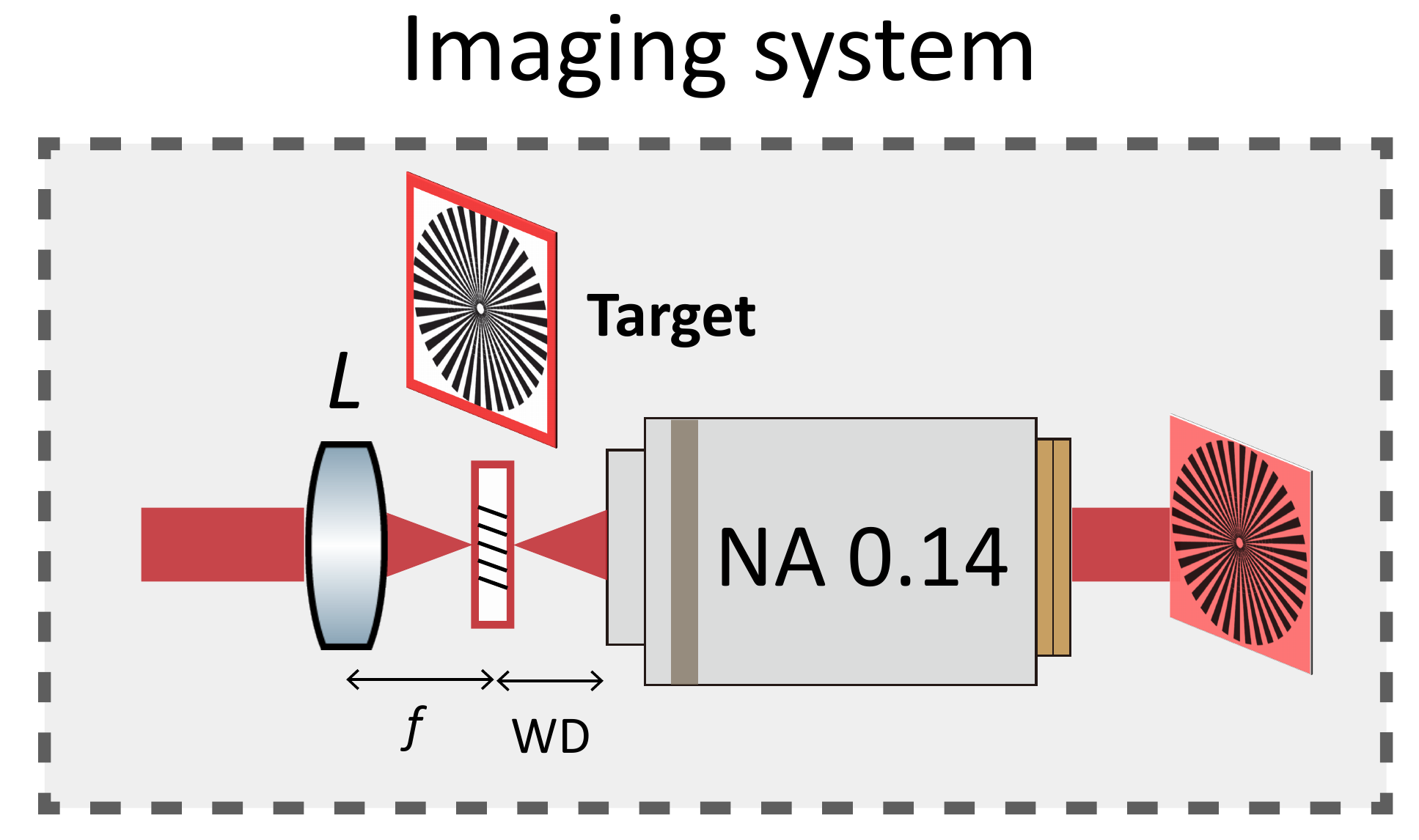}
\caption{Schematic of imaging system employed to encode real image of a Siemens star target in the IR signal beam (and subsequently in the SFG beam). The system consists of a lens (L) with \textit{f}= 100 mm and an objective lens with NA of 0.14, placed in a confocal configuration.}
\label{fig:imaging-system}
\end{figure}

Figure~\ref{fig:imaging-system} shows a schematic of the imaging system built to encode the information of a target in the IR signal beam. The imaging system consists of a Keplerian telescope that projects the real image of the target onto the metasurface plane as explined below. First, a resolution target (Thorlabs, R1L3S5P) is placed at the focal distance of a plano-convex lens, \textit{L} with \textit{f}= 100 mm. The transversal position of the target (across the \textit{xy}-plane) is controlled by a translation mount (Thorlabs, XYF1/M). After the target, a 5X objective lens with NA of 0.14 (Mitutoyo, Plan Apo NIR Infinity) is used to collect the signal beam. The objective lens is fixed at its working distance,\textit{WD} from the target which results in a confocal configuration. Through this imaging system, a collimated IR signal beam containing the real image of the target is obtained. The collimated signal beam is focused on the metasurface plane by a lens (see \textit{L1} in Figure~\ref{fig:setup}) projecting a real IR image of the target onto the metasurface plane.

\begin{figure}[htbp]
\centering
\includegraphics[width=1.0\linewidth]{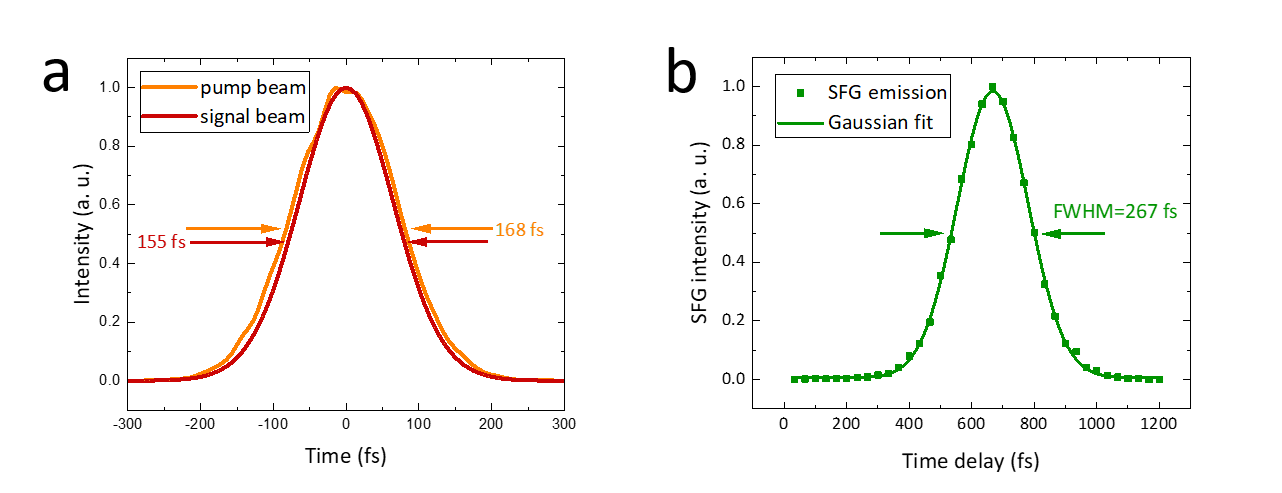}
\caption{Cross correlation of the pump pulsed laser beam with the signal pulsed laser beam. (a) The duration of the signal and pump laser beams was directly measured, giving a pulse-width of 155 and 168 fs, respectively. (b) The duration of the SFG emission was measured using an optical delay line, giving a pulse-width of 267 fs.}
\label{fig:pulse_duration}
\end{figure}

Figures~\ref{fig:pulse_duration}a and b show typical spectra of the Ti:Sapphire laser at 860 nm, and the OPO cavity at 1520 nm, performed with a spectral autocorrelator (Swamp Optics, Grenouille). The measurements were performed at the output of the OPO cavity (see Figure~\ref{fig:setup}), before the half-wave plates. Due to the femtosecond duration of the pulses, the dispersion effects introduced in the excitation beams when they travel through glass (focusing lens, half-wave plates, etc) are considered negligible. Thus, the duration of the pulses measured at the output of the OPO cavity correspond to the duration of the pulses at the metasurface plane. 

\subsubsection{SFG efficiency}

\begin{table}[h!]
\centering
\begin{tabular}{||c c c c c||} 
 \hline
 & P$^{ave}$(mW) & $\tau$ (fs) & P$^{peak}$(KW) & I(GW/cm$^2$) \\ [0.5ex] 
 \hline\hline
Pump beam & 16.4 & 168 & 1.235 & 0.78\\ 
Signal beam & 16.8 & 155 & 1.354 & 0.38\\ [1ex] 
 \hline
\end{tabular}
\caption{Experimental values of pump and signal beams employed in the calculation of the SFG efficiency.}
\label{table:1}
\end{table}

\begin{equation}
    \eta=P_{SFG}/P^{ave}_s=5 \times 10^{-8}
\end{equation}

\begin{equation}
    \eta_{norm} = \eta/P^{ave}_p=3.6 \times 10^{-6}~[W^{-1}]
\end{equation}

\end{document}